# The Wave Theory of the Field

by   [Walter E. R. Cassani](#)

---

## Summary

- **[Perturbations of the Schild's discrete space-time](#)**
  As a substitute for the current hypothesis of space-time continuity, we show the nature and the characteristics of a Schild discontinuous discrete space-time.
  The wave perturbations of its metrical structure are considered as perturbations of a new plausible and discrete discontinuous metrical "Ether".

- **[The wave hypothesis of the mass field](#)**
  Starting from the hypothesis of equality of the two energies: Einstein relativistic mass energy $E = m\,c^2$ and Planck radiation formula $E = h\,\nu$, we formulate the working hypothesis that all subatomic particles are elementary sources of spherical waves constituting on the whole the mass fields we attribute to the particles.

- **[The wave momentum](#)**
  Beginning from these «elementary waves» we discover a new law for the photon-particle elementary interaction coming from a simple symmetry principle.

- **[Energy and it's variation](#)**
  From the wave model, we causally deduce the photon's variational wave nature and, as a consequence, the relationship between «elementary waves» and the De Broglie waves.

- **[The Relative Symmetry Principle](#)**
  This simple and elementary symmetry principle is represented as the only law regulating the 4 interactions and unifying as a whole quantum mechanics and key phenomena of dynamics.

- **[A new wave interpretation of General Relativity](#)**
  An unexpected, experimental and descriptive completion of general relativity seems to have an inevitable causal connection with quantum mechanics, realizing the dream of unifying the relativistic and quantum physics that Einstein had pursued for a long time.

● **The wave description of Compton effect**
The explicative effectiveness of the new unification between quantum mechanics and general relativity is clearly shown by a wave interpretation of experimental data obtained by Compton effect, by applying the new wave laws on the base of the relative symmetry principle.

● **The wave model of Electron**
A further use of wave Compton effect leads us to discover an extraordinary mechanism of wave resonance which is able to verify the possible existence of a source of elementary waves called: particle.

● **The creation of wave pairs**
The generalization of the same mechanism of resonance for elementary waves allows us to justify the creation of particle-antiparticle pairs.

● **The wave interpretation of Lorentz force**
The appliance of a dynamic orientation for the same wave mechanism we identify with particles, shows its effects in presence of a magnetic field. It follows that Lorentz force is a consequence of the relativistic Doppler effect of the directed wave system constituting the particle and its field.

● **The wave nature of Inertia**
The wave nature of masses and the generalized effect of the relative symmetry principle leads us to consider the inertia as a natural and local consequence of the wave structure of bodies.

● **The wave nature of Gravity**
The same changeable model used for explaining inertia is also valid for describing a wave quantum gravitational interaction. The latter is a completely quantum interaction, but it is totally different from the "quantum gravity" we have till now considered.

● **Terminal Velocities for masses**
The exclusive wave nature of bodies and of the space-time quantization causes the impossibility for the masses in motion of overcoming the velocity of their elementary waves which travel at the velocity of light and prevent their wavelength from being reduced below the limit of the space quantum discrete length L imposed by quantization owing to Doppler effect.

- **The Antigravitational Fifth Interaction**
  Since it is impossible to reconsider a continuous space-time hypothesis, we can understand the impossibility of reducing to the infinity a wavelength describing the body's mass. Consequently, we can understand the existence of a fifth repulsive interaction acting in a more evident way at cosmological level among the maxi-masses and preventing the matter from being excessively dense, as Einstein had foreseen.

- **The almost banal certainty of Uncertainty Principle.**
  A deterministic description of the mythical uncertainty principle derives causally from the structure and organization of the waves constituting the particle as an elementary wave source of field.

---

To show the entity of the hypotheses that support
Quantum Mechanics and Relativity with respect those
of the Wave Theory of the Field to see the scheme:

## Expl-Hyph.pdf

---

To contact  Walter E. R. Cassani  e-mail :  waltercassani@tiscalinet.it

# Perturbations of a Schild discrete Space-time

Let us consider all the events of a Minkowski space-time in which the four coordinates **t, x, y, z,** are integers, and the velocity of light C is chosen as an unit. These events form a hyper cubic lattice.

The null lines joining the event points of the lattice derive from Lorentz transformations which make the lattice an invariant set. We will call those lines **"integral null lines"**, and the transformations **"Lorentz integral transformations".**

The contemporaneous time lines which are traced on the lines parallel to the axis of time **T** by Lorentz integral transformations, will be called time integral lines.

Let us construct, therefore, the model of this **Schild<sup>*</sup> discrete space-time** as a lattice made by discrete moduli of length **L**. The elementary discrete length **L** is considered as the smallest non-null space interval among the event points of the lattice.

The integral time lines are to their turn composed by moduli of discrete time **T** considered as the smallest non-null time interval among the event points of the lattice.

In the Schild discrete space-time, the dimensional quanta of time **T** can have a certain kind of dimensional elasticity, and their value can locally vary in any zone of the lattice, constituting a variation in the temporal uniformity of the **hyper cubic lattice**.

In this work, the space–time perturbations of the lattice structure are supposed to propagate in the lattice itself, while their motion is supposed to consist in a regular temporal sequence of points inside the lattice in order to bind the following points with null integral lines and temporal integral lines.

The propagation velocities of the perturbations of the lattice are the ratios between space and time quanta of the lattice, and they are associated with the integral Lorentz transformations described by the components **t,x,y,z,** of primitive integral vectors that must fulfill the diophantine equation:

$$1) \qquad ( t^2 - x^2 - y^2 - z^2) = 1$$

Thanks to it we attain the velocities:

$$2) \qquad v^2 = \left( \frac{x^2}{t^2} + \frac{y^2}{t^2} + \frac{z^2}{t^2} \right)$$

It follows that:

$$v = \frac{1}{t}\sqrt{t^2 - 1}$$

**3)**

If we consider the dimensional quantum **L** as invariant, we are forced to admit that the only variant quantum is the temporal quantum **T**. All of that involves an unequivocal choice of the possible values for the velocities associated with the integral time lines.

The time has two limit cases:

1) considering **t = 2**, the lowest possible velocity for a perturbation of the lattice is given by:

$$v = \frac{1}{2}\sqrt{3}c = c \cdot 0,866025$$

**4)**

2) for **t** tending to infinity, the maximum attainable velocity **V** tends to **C**.

In the lattice the first condition describes a perturbation of maximum local deformation of the lattice, which propagates from a zone of the lattice where is the temporal equivalent of a 3° hyperbolic space. This zone is made by points lying on the coherent sphere of maximum curvature, and its equation is **1)**.

This zone is subjected to the maximum dimensional variation in time quanta **T** which Schild says to be given by:

$$T = \sqrt{x^2 + y^2 + z^2} < x + y + z \leq \sqrt{3\left(x^2 + y^2 + z^2\right)} < 2T$$

**5)**

The second limit case describes the velocity of perturbation **V** tending to **C** for **t** tending to infinity, which moves in an almost plane lattice having a minimum curvature, and therefore a great but still not infinite radius.

**FIG.1**

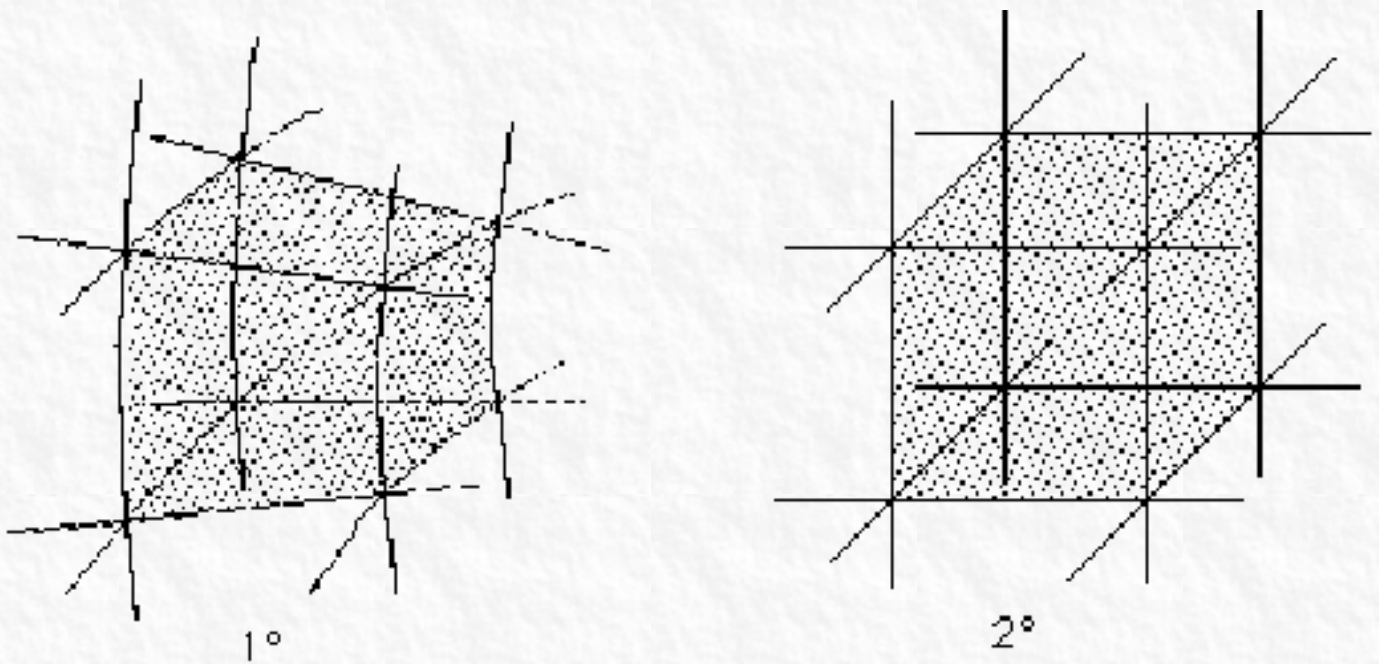

1°                                                          2°

In the first case, we consider the temporal perturbations of the lattice as geometrical variations of the moduli of length **L** which have, on the contrary, a constant length as in the second case. This length verifies the limit of the curvature condition for an infinite radius.

Perturbation fronts which behave like plane wave-fronts in the lattice, entering the field of curvature of spherical wave-fronts, move in the lattice according to the dimensional characteristics they meet in the spherical space-time established by the pre-existent spherical perturbations.

When the spherical perturbations are locally and periodically induced (we will see later a plausible model) a field of spherical waves develops and its curvature can influence the propagation direction of the perturbation surfaces introduced in the space- time of its existence.

**\***
*Alfred Schild* --Physical Review 1948. Vol. 73, pag. 414-415

*Alfred Schild* -- Canadian Journal of Mathematics 1948, pag. 29-47

---

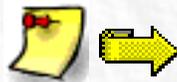

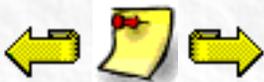

# The wave hypothesis of mass field

Now, we show the hypothesis that is at the basis of the Wave Field Theory.

**" Every force or potential field attributable to a particle endowed with a mass is a field of elementary waves, which has its own source within the particle."**

The behavior of waves and sources is determined by mutual interactions among waves, and between waves and sources. The particle's mass energy,:

$$6) \quad E_m = m\,c^2$$

is equal to Planck's wave energy attributable to the elementary waves constituting its field

$$7) \quad E_p = h\,\nu$$

It follows that:

$$8) \quad m\,c^2 = h\,\nu$$

Expressing the equality of the two energies in terms of wavelength:

$$9) \quad \lambda = \frac{h}{cm}$$

The particle having rest mass $m_0$, has a wavelength $\lambda_0$, which is equal to Compton wavelength.

$$10) \quad \lambda_0 = \frac{h}{cm_0}$$

The particle-wave source at rest has in any direction in its vicinity the same wavelength; the wave field has a spherical symmetry and for an ideal observer it emits elementary spherical waves.

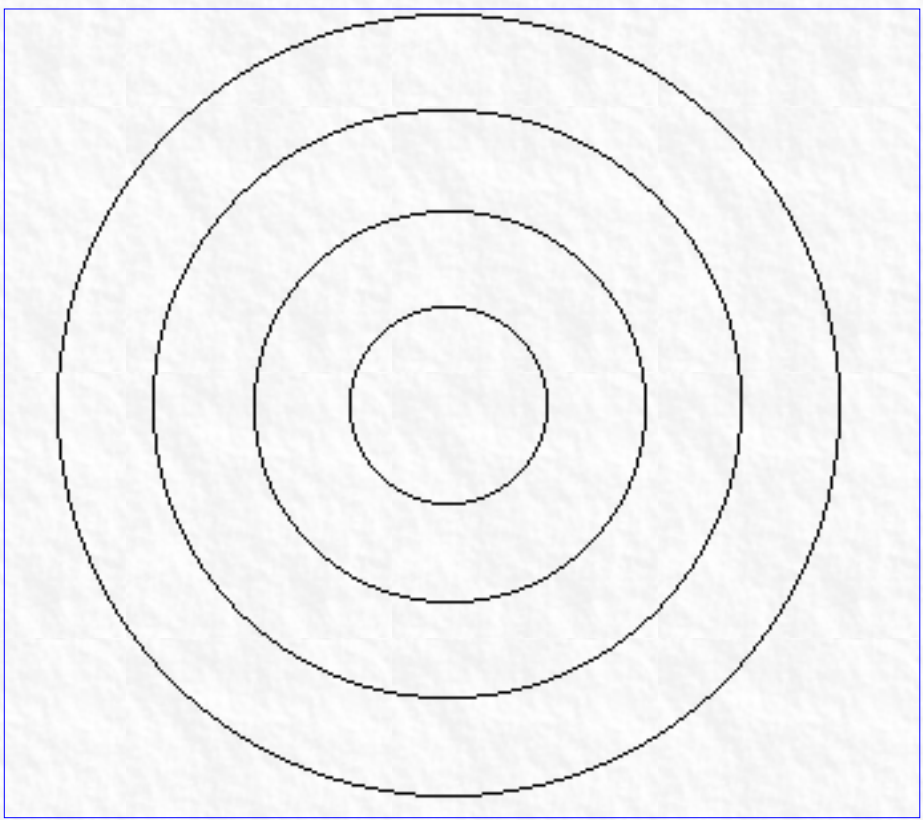

**FIG. 2**

An ideal observer who observes the motion of this particle in his frame of reference, verifies the existence of a wave number deriving from relativistic Doppler effect which enables him to represent the particle as in Fig. 3.

**FIG. 3**

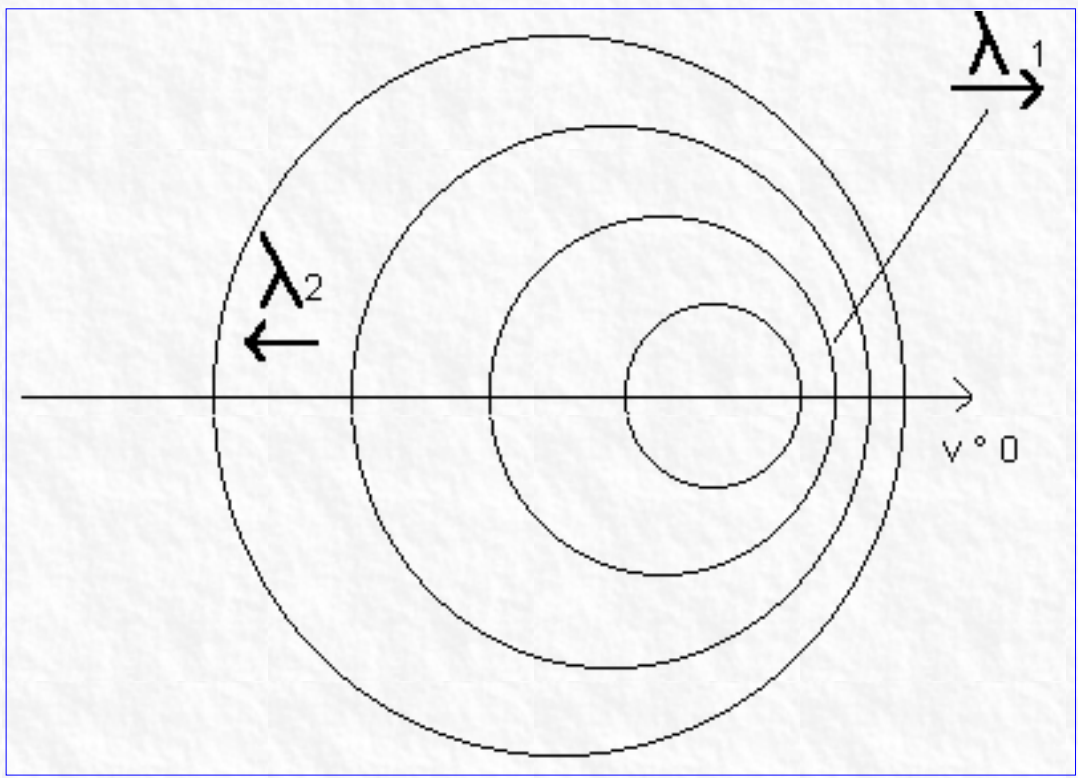

The ideal observer verifies a wavelength $\lambda_1 < \lambda_0$, that is shorter than that of the particle at rest, when the target angle compared to the particle's direction is Ø = 0.

$$\frac{1}{\lambda} = \frac{\left(1 + \dfrac{v}{c}\cos\phi\right)}{\lambda_0 \sqrt{1 - \dfrac{v^2}{c^2}}}$$

**11)**

When the angle is **Ø = π**, he observes a wavelength $\lambda_2 > \lambda_0$, that is greater than that of the particle at rest.
A real observer, unlike the ideal observer, verifies only some variations in the wave energy of elementary waves. When he observes a sudden acceleration of the particle, he can verify the existence of a wave train emission formed by a certain number of wave-fronts. The speeder the particle-wave source is, the shorter the wavelength of the wave-fronts is.

**FIG. 4**

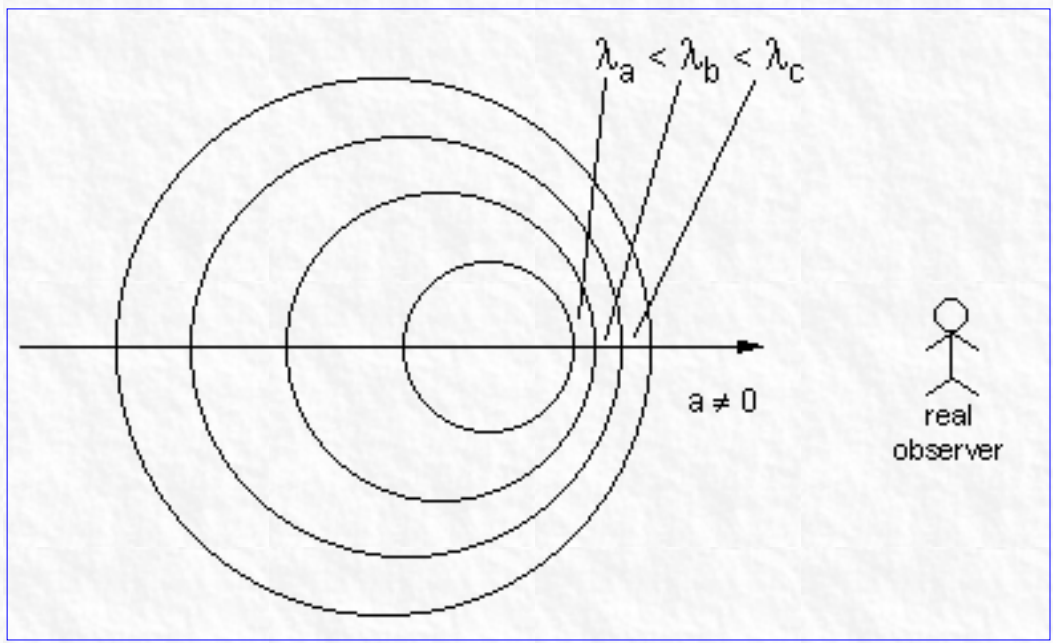

This wave train formed by waves of different wavelength will never be monochromatic and the real observer will call it "photon".

There are therefore two types of energy: the wave energy of elementary waves whose existence can be deduced but not perceived, and the energy formed by the variation in the energy of elementary waves, that we can verify through our senses and instruments and to which all of us react.

There is nothing mysterious in the hypothesis of the existence of the elementary waves and their energy. Nevertheless, we have to verify if the introduction of this hypothesis in physics is able to improve our comprehension of physical phenomena and of the theories supporting them.

With this view, let us admit that the symmetry properties of the field of the particle-wave source is an essential condition for the existence of the particle considered as an integral part of its energy and potential field.

Let us introduce a variation in wave energy in the vicinity of the particle considered as a perturbation of the state of wave symmetry of its field.

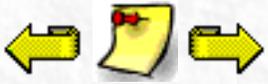

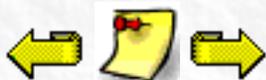

# The Wave Momentum

Let's observe in an ideal experiment an electron, considered as isolated in a space without fields, giving out some spherical elementary waves having the characteristic wavelength for the le rest electron. In **Fig.1** the energetic-wave distribution has all around a spherical symmetry.

**Fig. 4**

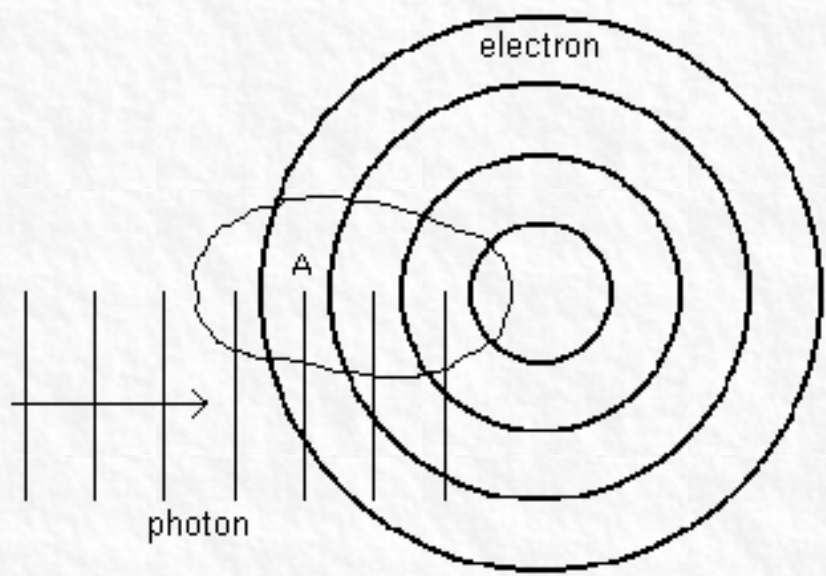

- From the infinite a plane wave train-energy photon $E_i = h \lambda_i$. approaches. The energy of the photon in the **A** zone is superimposed to the energy of the wave field of the electron (a part of the spherical waves can be considered as plane and parallel to the plane waves of the photon).

- In that zone **A** two energies coexist at the same time: $E_e + E_i$.

- The energetic distribution loses its **symmetry** all around the electron, when an anisotropic variation of energy happens around it.

Let's consider that a mechanism tending to restore the isotropy of the energetic variation starts at this moment.

- The electron-source of waves reacts moving on the opposite direction to the one where the previous variation of energy had happened.

- The motion of the electron varies the wavelength given out in the direction of the $\lambda_{e1}$, motion.

- It makes it shorter, while it lengthens the wavelength $\lambda_{e2}$, of the waves given out in the opposite direction to the motion, in compliance with the relativistic Doppler effect.
- When the particle is speed enough to allow the wave density, which is in front, to be equal to the density, which is at the back, the symmetry of the wave variation is reestablished and the wave state remains unchanged, until the arrival of another energetic variation.

It is possible to describe the wave-numbers situation of wave-source-particle and the wave-train-photon.

**12)**

$$\frac{1}{\lambda_1} = \frac{1}{\lambda_2} + \frac{1}{\lambda_i}$$

Here is represented the symmetry to which the disposition of the wave numbers around the particle has to tend:

**13)**

$$\frac{1}{\lambda_i} = \frac{1}{\lambda_1} - \frac{1}{\lambda_2}$$

Replacing the wave numbers of the particle with the equivalent values according to the relativistic Doppler effect:

**14)**

$$\frac{1}{\lambda_i} = \frac{2\dfrac{v}{c}}{\lambda_0 \sqrt{1 - \dfrac{v^2}{c^2}}}$$

In order to express the energetic content of the waves, we need to multiply both terms by **h**, (the energy quantum) and we obtain the relative relativistic momentum:

**15)**

$$\frac{h}{\lambda_i} = \frac{2h\upsilon}{\lambda_0 c} \frac{1}{\sqrt{1 - \dfrac{\upsilon^2}{c^2}}}$$

The photon transfers half of its momentum to the particle (in the following pages, in the "Wave analysis of the Compton effect", you will see this is only the first part of the diffraction phenomenon between photon and particle):

**16)**

$$p = \frac{2h\upsilon}{\lambda_0 c} \frac{1}{\sqrt{1 - \dfrac{\upsilon^2}{c^2}}}$$

**<span style="color:red">This is the real wave mechanism to transfer causally the energy of a photon to a particle.</span>**

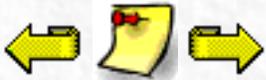

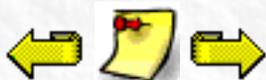

# Energy and its variations

The waves of every particle-source of waves that is in motion compared with an ideal observer, are seen by him as endowed with wavelengths having values that are analogous to the relativistic Doppler effect formula.

**Fig. 5**

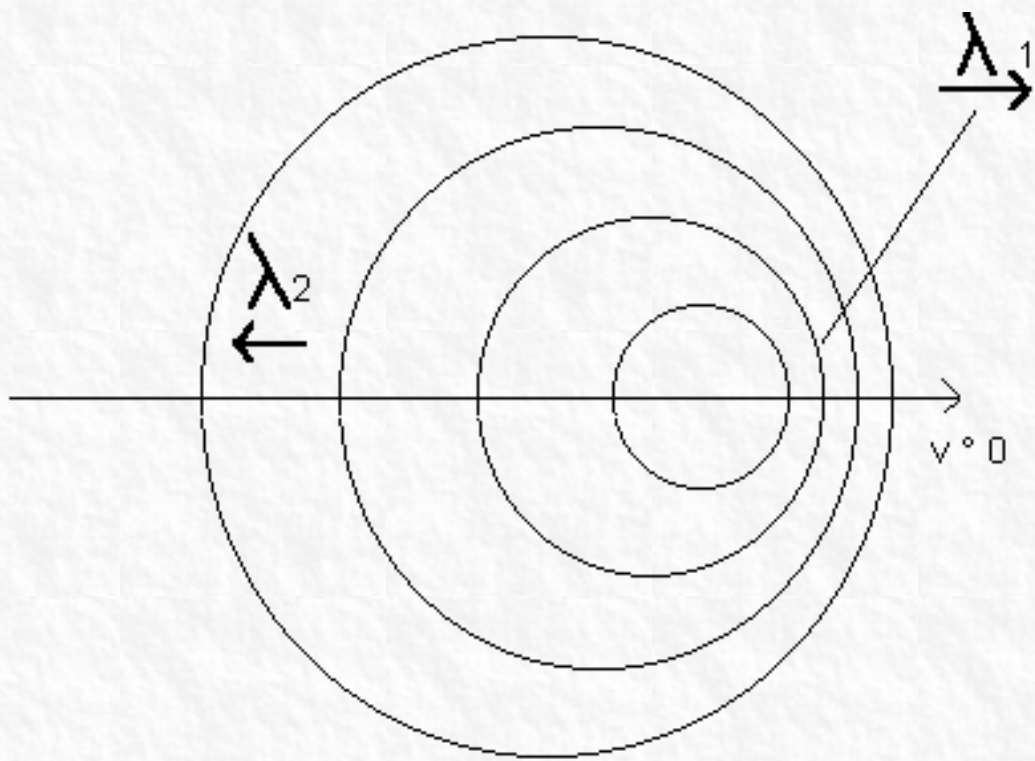

We mark with $E_1$ the energy of the waves in front of the particle in motion, and with $E_2$ the energy of the waves behind the particle, (that are different because of the Doppler effect). Firstly, the real observer sees the particle getting closer to him, after, passing in front of him and finally, going away. This experimented variation is $\Delta E$.

**18)**

$$\Delta E = E_1 - E_2$$

He verifies an energy variation deriving from the particle and describes the observable energy variation starting from the energy of the rest particle that cannot be observed:

$$E_0 = m_0\, c^2 = h_0\, \nu_0$$

The theory could be originally falsified with the verification of an experiment aiming at showing the possible perception of a photon composed by decreasing frequencies, and coming from a particle in an uniform motion, *(for example, setting a detector on a rotating drum, and reflecting this photon in a "mirror at optic coniugio of phase", that is a mirror endowed with a device that inverts the sequence of wave-fronts, compared with the normal reflection ).*
**<span style="color:red">This would be an original, experimental, Popperian attempt to falsify the new wave model.</span>**

Actually, if we consider the energy of the particle:

**Fig. 6**

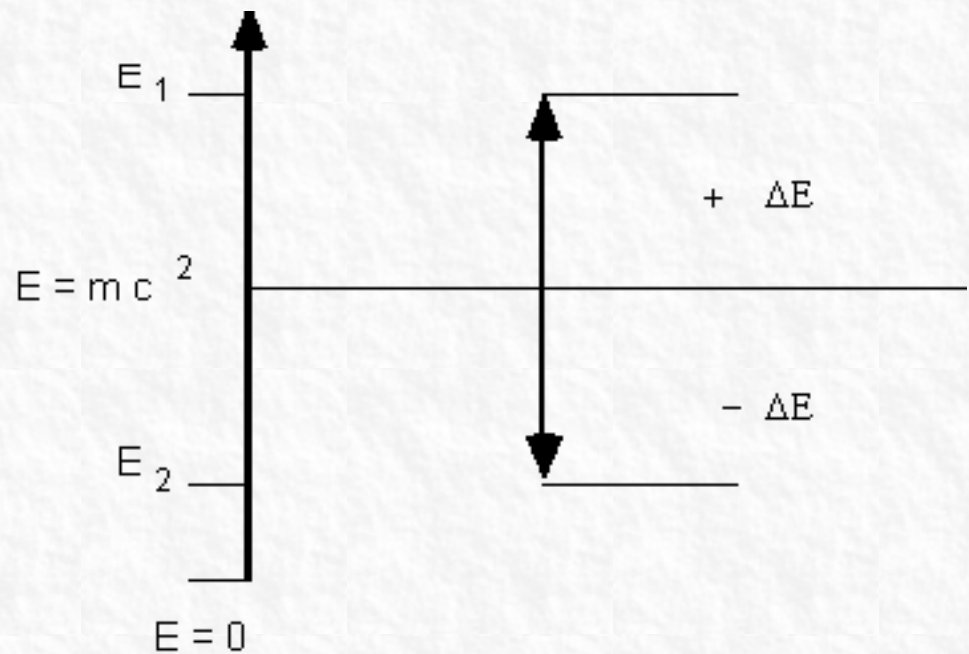

we see that the values of a perceivable energy oscillate around **$E = mc^2$**.

It follows that the average of the variation of the really observed wave energy is equivalent to the relativistic momentum of the particle:

**19)**

$$\frac{\Delta E_\lambda}{2c} = \frac{1}{2}\left|\frac{h}{\lambda_1} - \frac{h}{\lambda_2}\right| = \frac{h\upsilon}{\lambda_0 c}\frac{1}{\sqrt{1-\beta^2}} = p_\lambda = \frac{m_0\upsilon}{\sqrt{1-\beta^2}}$$

$p_\lambda$ is the same momentum described by **De Broglie** for the wave that is associated with the particle in motion having a wavelength:

**20)**

$$\lambda_B = \frac{\lambda_0 c}{\upsilon} \sqrt{1 - \beta^2}$$

In this way the energy variation is identified with the energy of the photon that can be perceived by the real observer:

21)

$$\Delta E = \frac{E_1 - E_2}{2} = cp$$

while the medium energy of the particle is its relativistic energy:

22)

$$E_m = \frac{E_1 - E_2}{2} = \frac{m_0 c^2}{\sqrt{1 - \beta^2}}$$

**So we have connected the wave dynamics, of the new wave model, with the De Broglie wave and with the relativistic dynamics.**

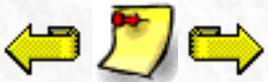

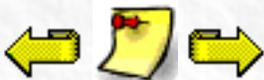

# The Relative Symmetry Principle

A new general physical principle can be enunciated after the considerations on symmetry which have led us to a purely geometric justification of momentum.

We will call this principle: " Relative Symmetry Principle "

It says that every body, being a source of elementary waves, behaves so as to maintain symmetrical every wave "variation" all around itself.

For the Relative Symmetry Principle, the variation of the wave energy is defined by the formula:

**23)**

$$\frac{dE_\lambda(\phi)}{dt} = \frac{dE_\lambda(\phi + \pi)}{dt}$$

This formula describes the variation of the wave energy that intervenes in the wave field of any mass, in the same way as the variation of the wave energy taking place in the symmetrical point, opposite to the barycenter.

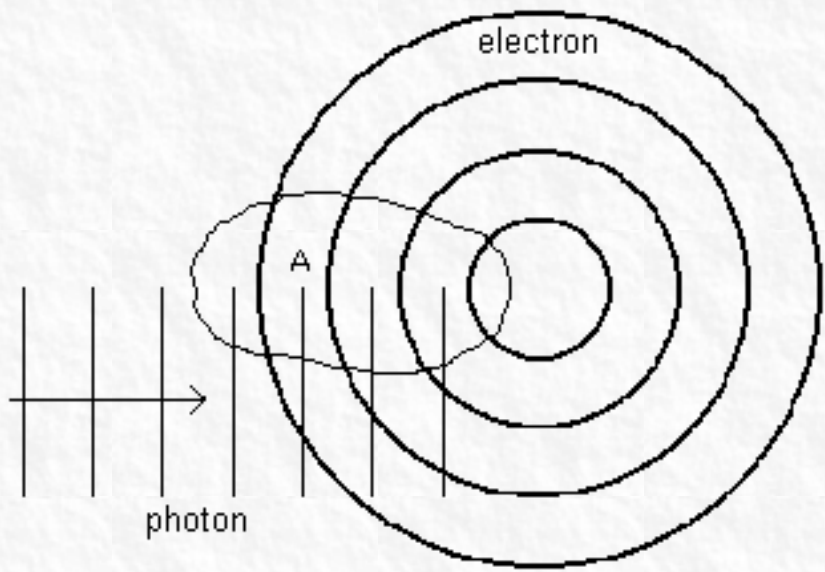

As indicated afterwards, this new principle will replace the Energy Conservation Principle and will be more descriptive and general than the first one.

Thanks to it, we will be able to justify the entity and the behavior of the four known interactions, and to foresee, indirectly, the existence of a fifth repulsive interaction.

*(The prediction and the behaviors of a fifth repulsive interaction have already been described in 1984 in the book: The Unified Field, already quoted above).*

We can know the explicative ability of this principle by deriving from the isotropy properties of the space-time a relativistic description of momentum, through which we naturally reach the relativistic variation of  mass.

The unified application of this principle to all the phenomena of interaction between radiation and matter, and of motion of bodies sets aside the observability conditions, and also acts where we are not able to directly observe it in its casual developments.

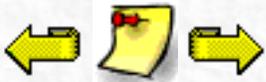

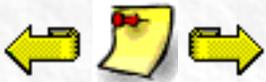

# The wave interpretation of General Relativity

The shift velocities of the temporal perturbations inside the hyper cubic lattice, in the Schild discrete space-time depend on the fields of curvature the perturbations meet during their propagation. This involves the wave interpretation of field in the explanation of the interaction of light with gravitational fields. We can introduce the wave interpretation of the formula describing the angle of deviation of light from the solar mass.

**47)**

$$\alpha = \frac{4Gm_s}{c^2 r_s}$$

Describing from the wave point of view the ms solar mass having $m_s$, as a radius $r_s$, it follows that:

**48)**

$$\alpha = \frac{4Gh}{c^3 \lambda_s r_s}$$

In the wave language the angle of deviation of the photon-wave train is proportional to the wave number of the deviant mass and inversely proportional to the radius of the mass waves interacting with the incident photons.

The plane waves of the photon having a great radius of curvature and little energy are influenced by the spherical waves system of the Sun, having a smaller radius of curvature and great energy . Besides the **G.R.** description, the deviation should be also proportional to the wavelength of the diverted photon, since there are two wave agents acting in the phenomenon: the waves of the photon having $\lambda_i$ as a wavelength and the mass waves of the Sun.

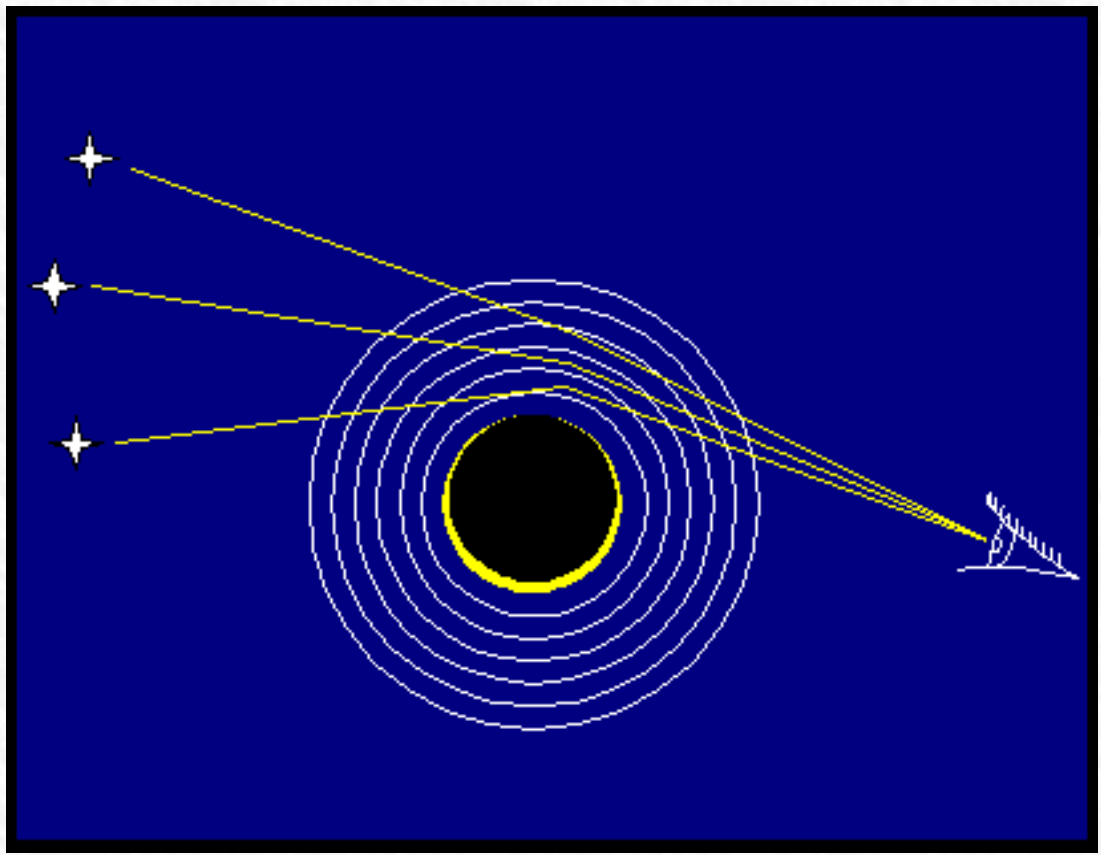

*(Till now, a dependence from the wavelength of photon has been neither verified nor considered; it depends on the exiguity of the angular deviation effect observable with macroscopic bodies, that in the case of the solar deviation would have increased the angle of deviation only by an insignificant umpteenth decimal).*

In order to understand the phenomenon in its possible variations, it is necessary to consider the role of the radius of deviation when the involved masses get smaller.

We can make a complete wave description of variables by adding a term to the **48**, and generalizing for any mass.

**49)**

$$\alpha = \frac{4Gh}{c^3 \lambda r} + \frac{\lambda_i}{r}$$

We can immediately notice that the added term ( $\lambda_i$ / **r** ) is not really significant in determining the value of the angle of deviation, when in the phenomenon great deviant masses having a great radius are involved. On the contrary, when we carry out the experimentations with smaller and smaller masses having shorter and shorter distances from the center of the wave field, the first term loses importance while the added one becomes predominant.

When we experiment the effect of the deviation of light from the matter, we immediately meet with a macroscopic deviation effect that in Optics is called Diffraction.

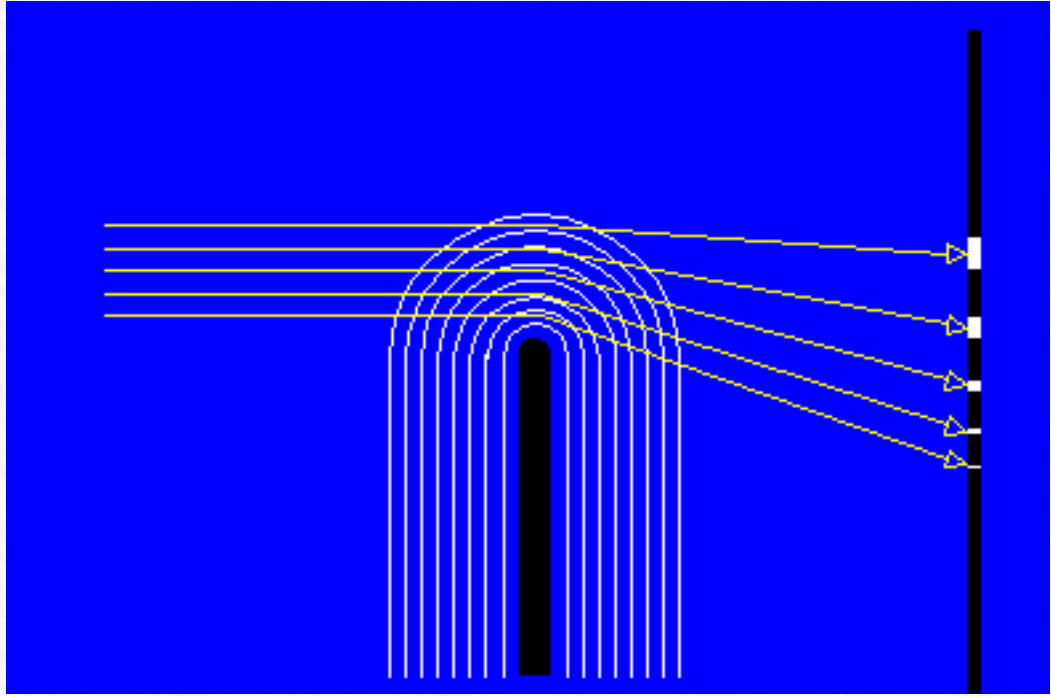

The optical diffraction from a thin edge is an example of deviation of light that on a small scale can be compared to the solar deviation.

According to the wave hypothesis of field, in the diffraction from a thin edge, the obstacle placed on the path of the waves of the photon traveling in the light ray gives out a series of wave-fronts, that create all around the profile of the obstacle a field of waves similar to the obstacle in its shape.

This field of waves derives from the spherical waves given out by all the elementary particles constituting the matter of the obstacle at a subatomic level.
If the obstacle is formed by the profile of an upright blade, it gives a wave of semi cylinder shape.

Being nearby the edge, the experimental wave situation is, on an inferior scale, quite similar to the one verifying in the deviation of light from the solar mass; but, in the last one there are some quantitative differences underlining the validity of the qualitative considerations on the importance of the smallness of the radius of the deviant waves.

The angle of deviation of light imposed in the diffraction phenomenon from an edge is, in optics, more marked than in the phenomenon described by the general relativity. There are mainly two changing data:

- ❍ The frequency given out by the edge of the deviant obstacle is much smaller than the one given out by the

Sun. Therefore, its wavelength is much greater.

❍ The radius of the waves given out from the edge with which the light comes into contact during its passage near it, is much smaller than the radius of the spherical waves given out by the Sun, which start from its surface and have at least the same radius as the Sun itself.

The first conclusion we can draw is that, even if the size of the angle of deviation of light imposed by the general relativity seemed to depend on the mass value of the deviant body, the radius of the mass field is much more determining since it is the deviant cause.

Therefore, we suppose that the formulas of the general relativity must be considered only as first approximations of the true law, in which the really determining factor is the smallness of the radius of the mass waves involved in the diffraction phenomenon.

Yet, even this phenomenon gives only qualitative confirmations. In order to find a quantitative analysis, it is necessary to examine other phenomena of interaction between matter and light, in which the radius of the mass waves is still small, and it is even more determining in putting in evidence the action of the form waves in the orbit of the photons of light.

**The determining and definitive phenomenon happens in the adverse field: the Compton effect.**

If this was true, it would mean that it is possible to lead the General Relativity within the interactions of the radiation with the elementary particles, by using the wave description of field for the wave interpretation of the key phenomena of the Quantum Mechanics.

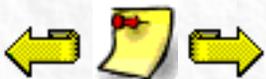

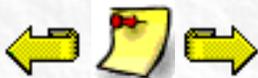

# The wave description of the Compton effect

We can verify the hypotheses and the laws deriving from the new wave theory, by using the Relative Symmetry Principle and the wave changes in the General Relativity, for a new purely wave interpretation of the simplest phenomenon of interaction between matter, in the form of elementary particle (electron), and the most elementary radiation (photon) in the Compton effect.

Let's observe, in four times, an ideal experiment of Compton interaction with a photon having $\lambda_i$ as a wavelength

$\lambda_i = 1 \cdot 10^{-10}$ **m** diverted by **90°** by a free electron, in a space deprived of significant fields.

1) the photon gets closer to the source of the field of the electron, and its energy is added to the one of the waves of the field of the electron in the **A** zone:

**50)**

$$E_i + E_e = \left(\frac{hc}{\lambda_i}\right) + \left(\frac{hc}{\lambda_e}\right)$$

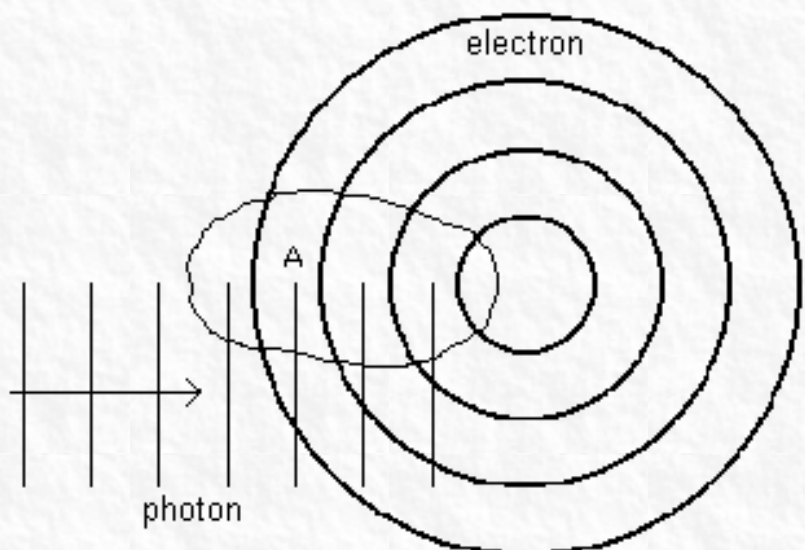

**Fig. 11**

Because of the wave variation, the Relative Symmetry Principle obliges the electron to the motion, imposing the $v_1$, velocity on it.

from **15)** we draw the $v_1$, velocity. It follows that:

**51)**

$$v_1 = \frac{1}{\sqrt{\dfrac{4\lambda_i^2 + \lambda_0^2}{\lambda_0^2 c^2}}}$$

Let's consider the frame of reference of the electron-source of waves that moves with the $v_1$ velocity. Observing the incident photon, we notice a decay due to Doppler effect, as resulting from the following formula, where the angle of incidence $\emptyset$ of the photon compared to the direction of the velocity of the electron is $\pi$.

**52)**

$$\lambda_{il} = \frac{\lambda_1 \sqrt{1 - \dfrac{v_1^2}{c^2}}}{\left(1 + \dfrac{v_1}{c}\cos\phi\right)}$$

After the decay, the photon is diffracted $\alpha = \pi/2$, by the spherical field of waves of the electron, in compliance with the **G.R.** modified formula, considering the first part insignificant.

**53)**

$$\alpha = \frac{\lambda_{il}}{r}$$

$$\alpha = \frac{\pi}{2} \quad r = \frac{2\lambda_{il}}{\pi}$$

for

The **r** radius is the least distance from the center of the particle-source of spherical waves, which is crossed by the incident photon before being diffracted. This radius plays a decisive role in the whole following study, introducing the descriptive possibilities of the theory through the new wave interpretation of the General Relativity.

Before that, a phenomenon of pure and simple deviation of the plane wave train-photon takes place from the wave field of the electron. Since the electron hasn't had time to accelerate (as we have seen for the solar deviation), it limits itself to the diffraction of the photon, maintaining its energy unchanged. The phenomenon is actually experimentally observed when the incident light ray is formed by several photons. It gets the presence of a component of the diverted radiation, with the same angle, which has the same energy of the incident radiation, and isn't therefore subject to decay.

As shown by the following formula drawn by the $v_1$, velocity and from $I_{i1}$ where the variables of the phenomenon play their role:

**54)**

$$\alpha = \frac{2\lambda_i^2}{2\sqrt{4\lambda_i^2 + \lambda_0^2}} \frac{1}{1 - \frac{\lambda_0 \cos\phi}{\sqrt{4\lambda_i^2 + \lambda_0^2}}}$$

Considering the wavelength of the incident photon and the wavelength of the deviant particle, we can foresee the maximum angle of deviation of the photon, (and this has an experimental importance), and know the least radius of the curve described by the photon all around the center of the wave field of the diffracting particle.

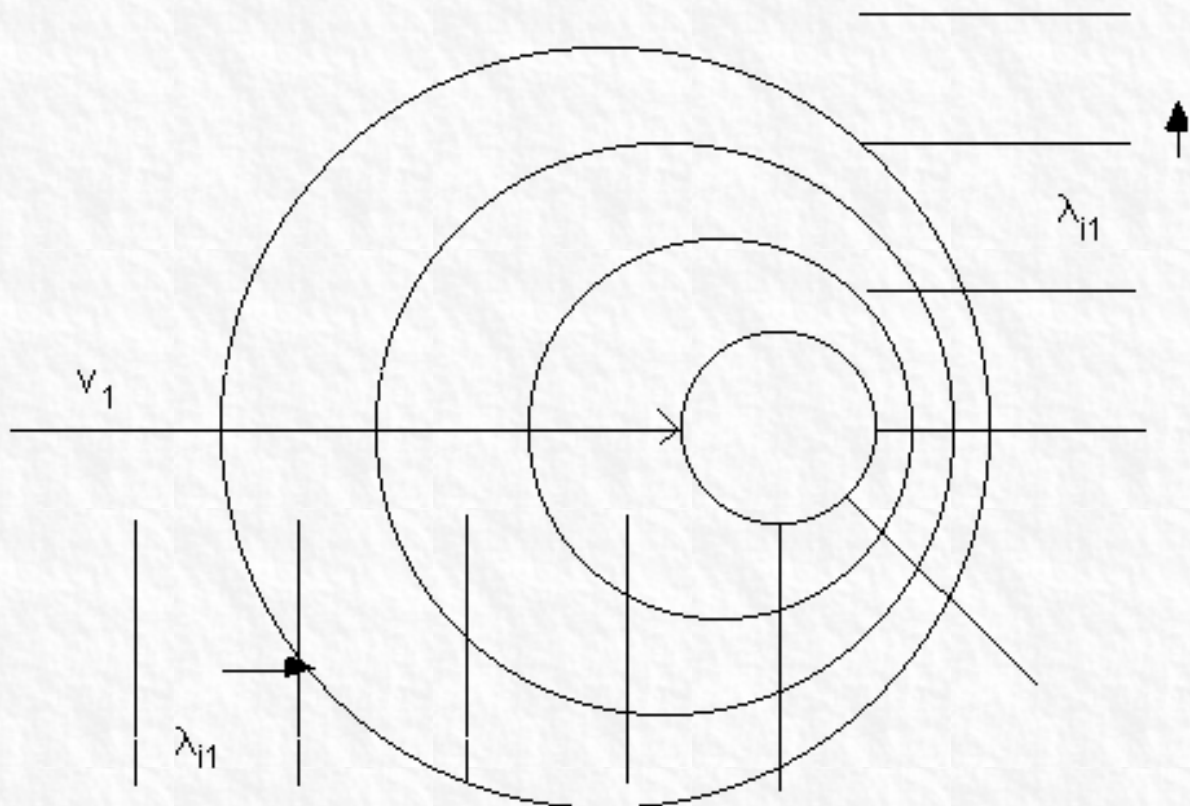

**Fig. 12**

Because of the acquired velocity, the field of waves of the electron is modified by the transverse relativistic effect. So, after undergoing a first decay, the photon triggers a new push effect, in compliance with the **Relative Symmetry Principle** (*), and develops a new momentum in the electron (where the transverse mass of the electron enters) having a direction that is orthogonal compared to the previous one.

(*) That since now will be called **R.S.P.**

The transverse relativistic wavelength of the electron (that justifies the existence of the transverse mass of Relativity) is:

**55)**

$$\lambda_{e1} = \frac{\lambda_e \sqrt{1 - \dfrac{v_1^2}{c^2}}}{\left(1 + \dfrac{v_1}{c}\cos\phi\right)} = \lambda_e \sqrt{1 - \frac{v_1^2}{c^2}}$$

If it is introduced in the R.S.P. formula, because of the push given by the photon having $\lambda_{i1}$, as a wavelength, it allows the electron to have a $v_2$ velocity that is orthogonal compared to the previous one:

56)

$$\frac{h}{\lambda_{i1}} = \frac{2hv_2}{\lambda_{e1}c} \frac{1}{\sqrt{1 - \dfrac{v_2^2}{c^2}}}$$

57)

$$v_2 = \frac{1}{\sqrt{\dfrac{4\lambda_{i1}^2 + \lambda_{e1}^2}{\lambda_{e1}^2 c^2}}}$$

Compared to the electron-source of waves, which moves with $v_2$, velocity, the diffracted photon further decays because of the Doppler effect, as given out by the electron, and its final wavelength becomes:

58)

$$\lambda_{i2} = \frac{\lambda_{i1} \sqrt{1 - \dfrac{v_2^2}{c^2}}}{\left(1 + \dfrac{v_2}{c}\cos\phi\right)}$$

The diffracted photon has as a whole:

59)

$$\Delta\lambda_i = \lambda_{i2} - \lambda_{i1}$$

while the electron has reached the velocity: $v_e = v_1 + v_2$ (*)

*(\*) The sum of these two velocities is not a simple vector sum. Lyevelin Thomas (\* \*) showed that a Lorentz transformation with $v_1$ velocity, followed by a second transformation with $v_2$ velocity, in another direction, doesn't lead to the same inertial reference reached by a single Lorentz transformation having $v_1 + v_2$. as a velocity.*

(\* \*) (H. A. Kramers, *Quantum Mechanics,* New York 1957)

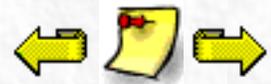

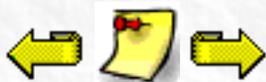

# The wave model of Electron

As explained above, the term added to the **G.R.** can become more and more determining as we experiment with deviations of photons with masses giving out spherical fields of waves composed by waves that are still close to their source of elementary waves.

These waves still have a very small radius, and therefore can interact with light, determining the greatest possible angle of deviation for any photons crossing the geometry of the space-time established by their presence.

**This happens when the relationship between the wavelength of the incident photon and the radius of the resonance orbit crossed by a wave, approaches the unit. In such a case, the first part of the general relativity formula <span style="color:red">loses importance</span> and can be totally neglected, while <span style="color:red">the added term becomes predominant</span>:**

**60)**

$$\alpha = \frac{\lambda_i}{r}$$

Let's consider that the angle of deviation of the photon is greater than the angles normally experimented in the Compton effect or that it is at least: $\alpha = 2\pi$.

Let's explain how can we reach such a deviation after passing through several wave situations, by using an electron having $\lambda_e$, as a wavelength.

This electron is invested by a photon having $\lambda_i = \lambda_e/2$, as a wavelength diverting it by an angle $\alpha = 2\pi$:

1) The Relative Symmetry Principle pushes the electron to the $v_1$, velocity, since the photon transfers half of its momentum to it. The photon, running after the electron, loses half of its momentum because of the Doppler effect, and its wavelength becomes $\lambda_{i1} = \lambda_e$.

2) Let's consider that the resultant photon is continuously diffracted by the field of spherical waves of the electron, completing another deviation equals to **180°**, returning to its original position.

3) The photon-wave train rotates on a circular orbit in a closed circuit, round the source of the field of spherical waves of the electron, completing an angle of **360°**.

4) if the length of the closed circuit crossed by the photon is equal to the final wavelength of the photon, the wave train is subject to the waves resonance law, which perpetuates the motion of the wave <span style="color:red">on the resonance orbit:</span>

**61)**

$$2\pi\, r_0 = n\,\lambda_0$$

5) When **n = 1**, the wave train, that has already completed a revolution, can complete more resolutions, superimposing all the wave-fronts to the first one, that had already placed itself in a resonance condition, on the orbit having $r_0$ as a radius.

Considering the whole wave-front, which tries to place itself in a resonance condition, only the part that is closest to the center of the deviant spherical wave field is able to place itself with one or more elementary surfaces, in the resonance condition.

The rest of the $L^2$ elementary wave-fronts constituting the wave-front is propagated on the paths depending on conditions of curvature imposed by the spherical waves of the field of wave belonging to the obstacle particle, that is an electron or a proton.

The diffracted wave-front is deformed, during its propagation, in the Schild space-time, and places itself according to a characteristic surface known as: Spherical Involvent or Spherical Involute.

The points intersected by the Spherical Involvent on the plane containing the resonance orbit enable us to describe the plane Involute curve that, according to a **x y** frame of reference having its origin in the center of the resonance orbit, and to a $x_1$, $y_1$, frame of reference which is parallel to the first one and rotates with the c velocity on the orbit, can be described by:

62)

$$x = r_0\,[(\omega + 2k\pi)\,\sin\omega + \cos\omega\,]$$

63)

$$y = r_0\,[\sin\omega - (\omega + 2k\pi)\,\cos\omega\,]$$

for $\quad k \in Z_0^+$

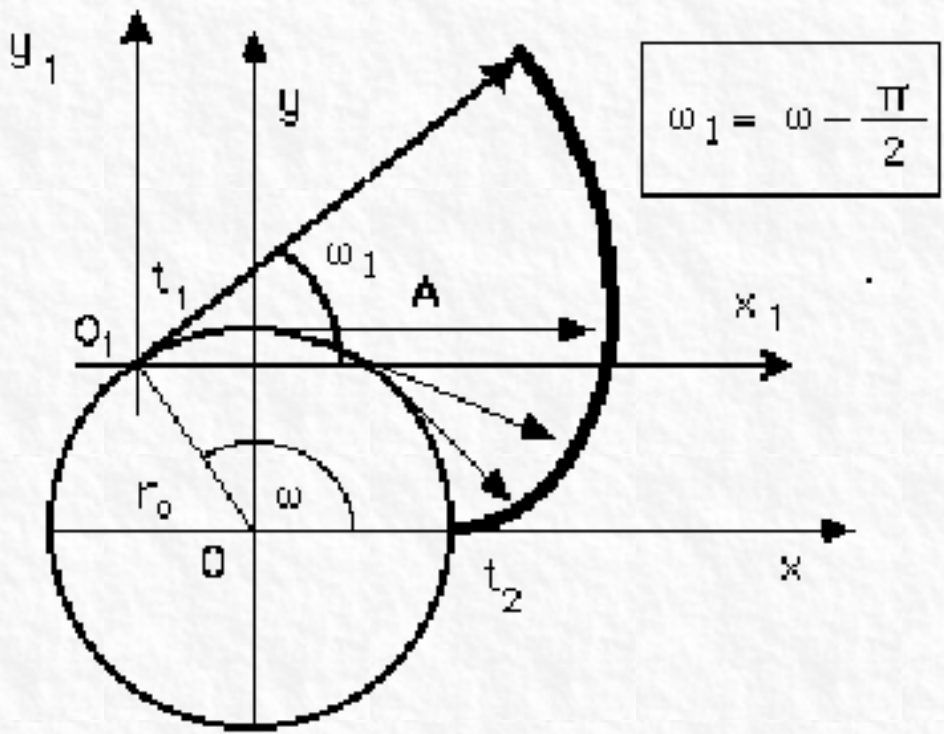

$$\omega_1 = \omega - \frac{\pi}{2}$$

**Fig. 13**

On the resonance orbit identified by the **Involvent**, an **A** vector rotates with an increasing **c** velocity, going from the $t_2$ to the $t_1$ time describing the **E** curve appeared in the $t_1$ time. The **Involvent** produces wave-fronts that are more and more approximately circular to the distances compared to which r becomes a negligible quantity, and it maintains the $\lambda_0$ wavelength constant.

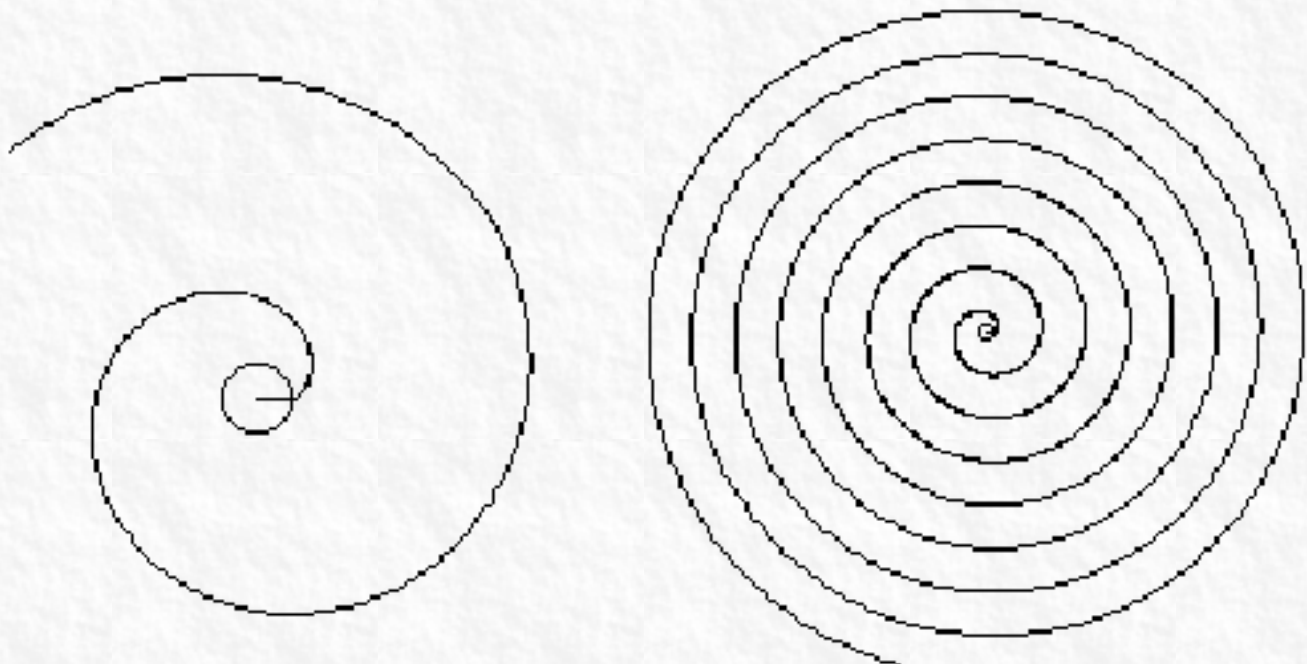

**Fig. 14**

It can also be bidimensionally described in an exponential form by the **A** vector given that:

**64)**

$$x = x_1 + r_0 \cos \omega \qquad y = y_1 + r_0 \sin \omega$$

**65)**

$$A = \left( r_0 \omega_1 \right) \cdot e^{i\left( \omega_1 + 2k\pi \right)}$$

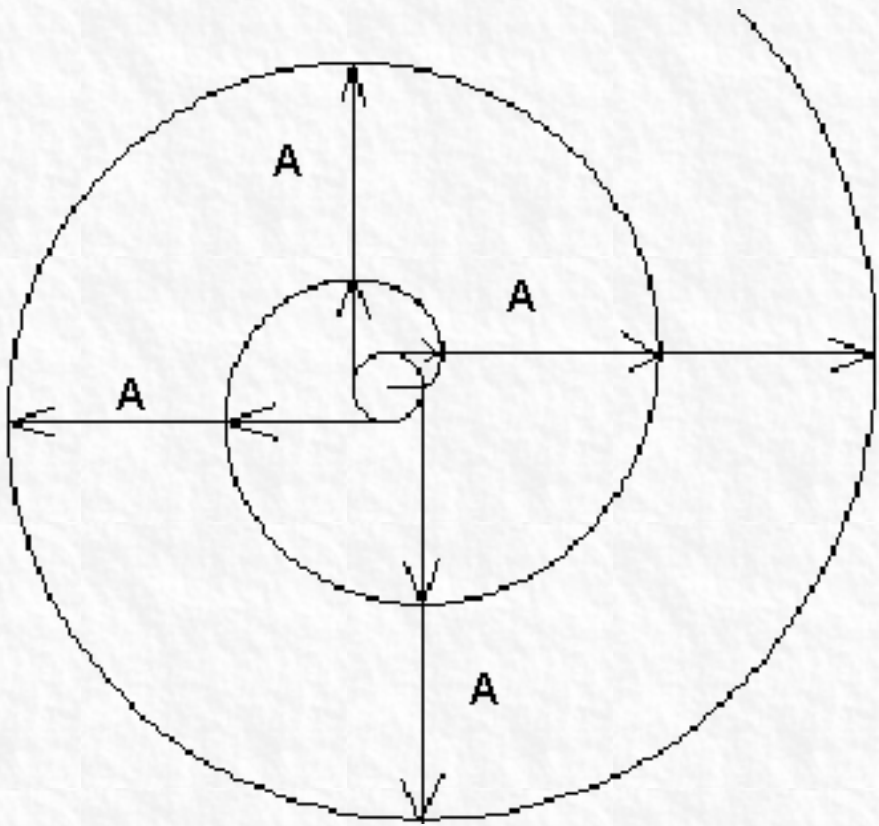

**Fig. 15**

We now can tridimensionally deal with the developments of the model, by ideally projecting in the **z** axis of the resonance orbit the orbit itself, so as to obtain a half positive and half negative infinite ideal cylinder.

Z

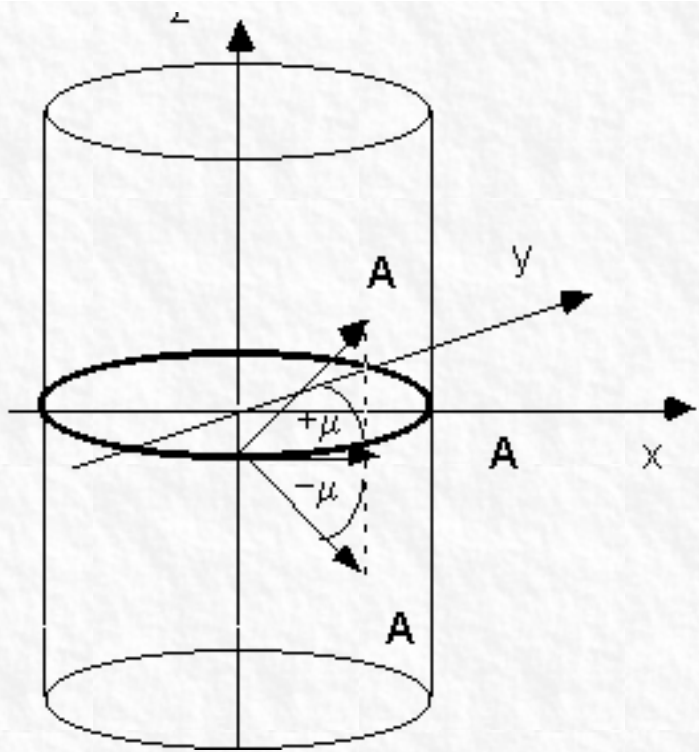

**Fig. 16**

Let's consider the set of the **A** vectors for a fan-shaped direction variation until reaching the ±μ angle.

**66)**

$$-\pi/4 \geq \mu \geq +\pi/4$$

In order to make the surface of the Involute spherical we displace the **A** vector, according to the temporal properties of the Schild lattice.
The values of this vector, in a Riemannian cylindrical geometry coincide with the surface of the cylinder and have at least a **2**μ.

Given that $\lambda_0 = 2\pi r_0$, we are able to describe the helicoid as a geodesic, that starting from the resonance orbit twines round the ideal cylinder, which is constituted by the projection of the resonance orbit.

Its modulus is given by:

**67)**

$$\mathbf{A} = \frac{r_0 \omega_1}{\cos \mu}$$

There are two helicoids: one for μ that develops from a point of the resonance orbit towards the positive **z**; the other, that is the mirror image of the first one, develops for μ that is propagated towards the negative **z**. For μ **= ±π/4** we have:

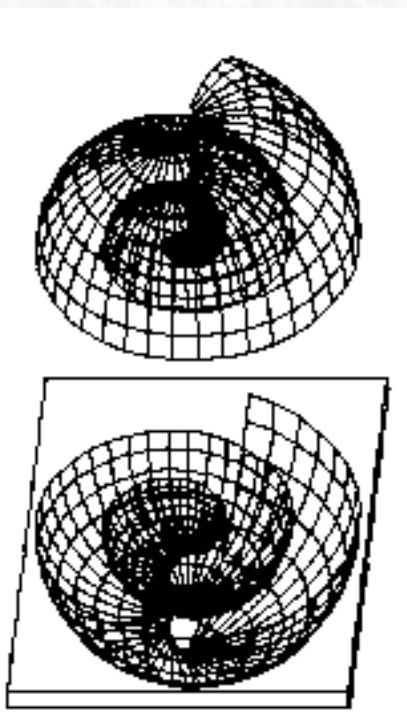

**Fig. 17**

**68)**

$$x = r_0 \cos \omega_1 \qquad y = r_0 \sin \omega_1 \qquad \pm z = r_0 w_1 \tan \mu$$

The spherical link of the helicoids with the involvent curve constitutes a surface that, during its evolution, forms the Spherical Involvent.

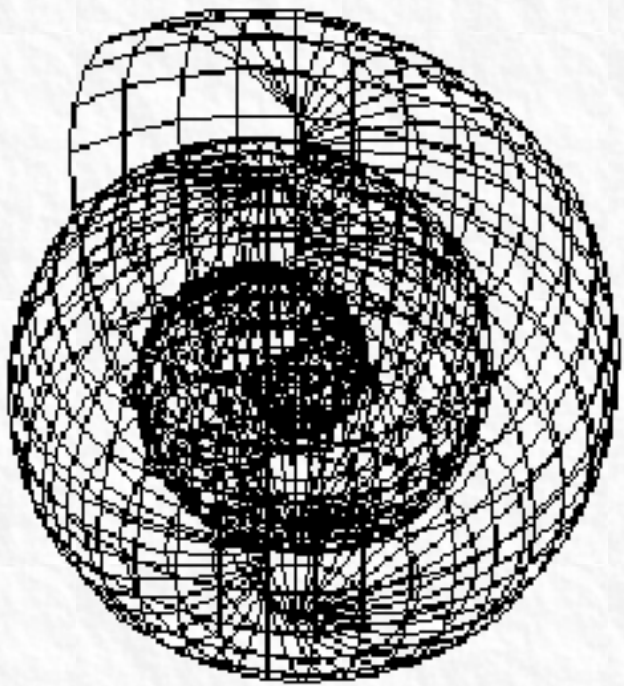

**Fig. 18**

<span style="color:red">**This is the**</span> **Spherical Involvent** <span style="color:red">**we identify with an elementary particle-rest source of**</span>

**waves, having  $m_0$  as a mass.**

**69)**

$$A = \frac{\left(\dfrac{\lambda_0}{2\pi}\,\omega_1\right) \cdot e^{i(\omega_1 + 2k\pi)}}{\cos\mu} \cdot \frac{\sqrt{1 - \dfrac{v^2}{c^2}}}{\left(1 + \dfrac{v}{c}\cos\phi\right)}$$

## The formula show the secret of the material micro-universe.

### The wave-particle.

As shown in the figure, the meridians appear perpendicular to the plane of the resonance orbit, but in reality (we'll explain it when we speak about the justification of charges), they are inclined and  like two screw threads they twine round the spherical surface.

A simple Basic program show the structure of the wave-particle: **Basic-involvent.htm**

The first one on the right-hand side and the other one, its mirror image, on the left-hand one *(the parallels and meridians don't exist but at the level of the elementary dimensions of the Schild space-time).*

We now have the data we can use to describe the Spherical Involvent and its dynamics, so as to establish a connection between the behavior of the particle-source of waves and the relativistic variation of the waves of the Spherical Involvent and identify it with the particle itself.

---

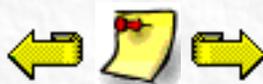

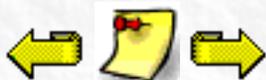

# The Creation of Wave Pairs

At the beginning of the paragraph concerning the creation of the plane Involute, we have considered a wave train that is diffracted by the field of an electron in the same way as in Compton effect, and has the wave-front arranged only by one side with respect to the electron's wave field.
Let us consider indeed a wave train that invests the electron frontally, and that is divided into two parts: one part of the wave-front is diffracted to the right, and the other one to the left.

**The final products are "two" spherical involutes– wave sources, both endowed with opposite Spins, the one that is the mirror image of the other.**

**The particle and the antiparticle.**

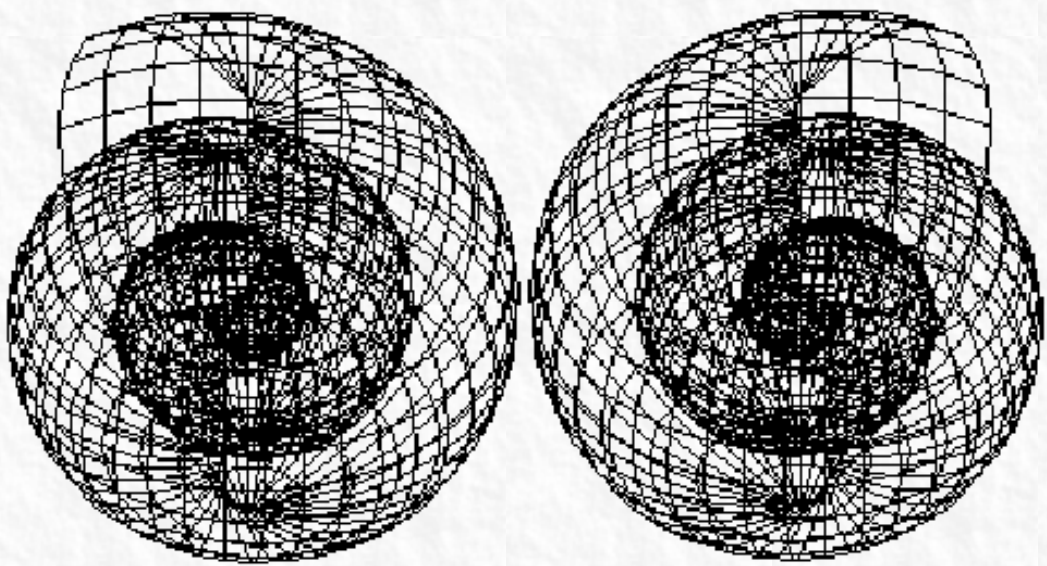

**FIG.19**

Is this therefore the phenomenon of the creation of pairs?

If the wavelength of incident photon is $\lambda_i = \lambda_e/\mathbf{2}$ and the diffracting obstacle is a free electron, we will have the production of an electron and a Dalitz pair formed by a positron and an electron.

If these two wave structures together with their creative obstacle are subjected to a magnetic field, they will form the three classical traces in a bubble chamber.

Now, if we subject the propagation of the waves creating the spherical involutes to relativistic conditions, we will have the possibility to justify the **Lorentz force** from a wave viewpoint, without invoking nor presupposing the existence of the electromagnetic theory at all.

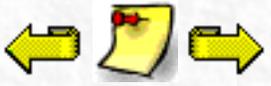

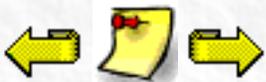

# The wave interpretation of the Lorentz force

In the phenomenon really experimented in bubble chamber, the effect of the Lorentz force diverge in opposite direction the positron and electron traces.

According to the Wave Field Theory, the Lorentz force bends the trajectory of an electron when its spin is parallel to the lines of force of a magnetic field, and its velocity develops on a plane orthogonal to the spin.

In the wave language, this condition imposes the spherical involute to move in a plane passing through its resonance orbit.

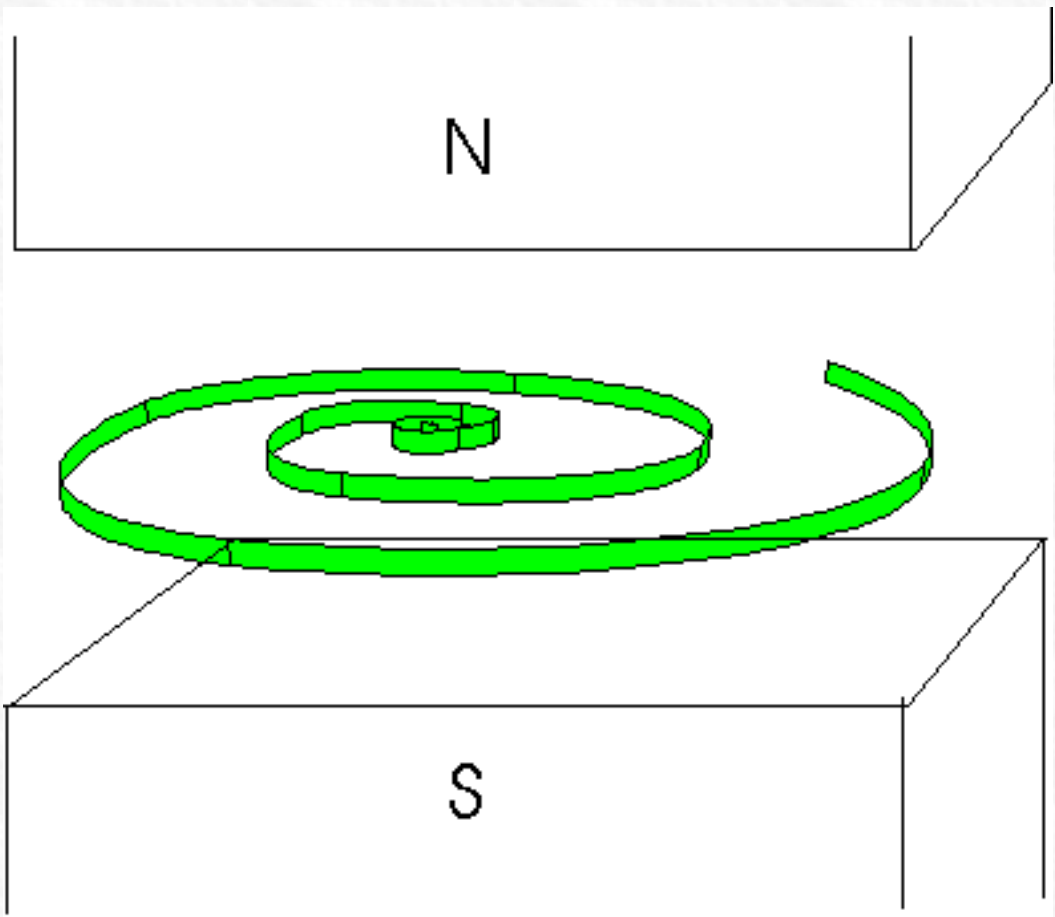

When this condition appears, the wave-front is in different times in four physically different conditions:

**1)** the wave-front, which moves in the resonance orbit and creates the involute, during its rotation proceeds firstly with velocity v in the same direction as the velocity of the center of the resonance orbit; therefore its wavelength emitted in this direction gets shorter owing to Doppler effect;

**2)** secondly, it moves in a direction transversal to velocity, leaving its wavelength unchanged;

**3)** thirdly, it moves in a direction opposite to velocity, increasing the wavelength owing to opposite Doppler effect;

**4)** finally, it moves in a direction transversal and opposite to the previous one, leaving the wavelength unchanged and returning to the first condition in order to repeat the whole cycle again.

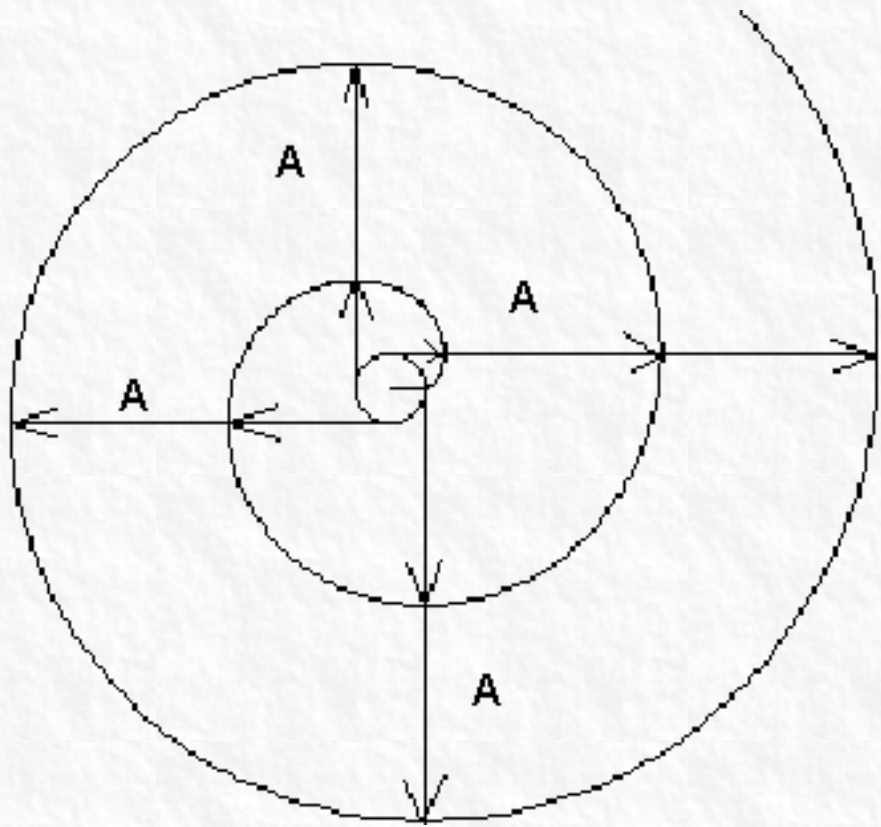

During the whole rotation, the resonant wave-front, that is subjected to relativistic Doppler effect, produces a deformed involute with an accumulation of wave-fronts by the side of the spherical involute, causing on the plane an asymmetrical variation in the wavelength of the waves emitted orthogonally with respect to velocity.

**70)**

$$q_0 \left( E + v \wedge B \right) = \frac{d}{dt} \frac{\left( \dfrac{h}{\lambda} \dfrac{v}{c} \right)}{\sqrt{1 - \dfrac{v^2}{c^2}}}$$

$q_0$ is the elementary charge, **E** is the electric field pushing at velocity **v** the electron which crosses the magnetic field **B** with the spin parallel to its lines of force.

Due to the relative symmetry principle, this causes a thrust in the opposite direction and bends the trajectory of the spherical involute in a direction and that of its mirror image in the opposite direction.

An electron, that is not subjected to a magnetic field, travels in straight line because its spin without having a constraint on the orientation is placed in the direction of motion. This is its smallest energy position and the greatest symmetry condition for an asymmetrical structure like the spherical involvent.

This explains the curl traces of charged particles in bubble chamber that losing velocity are in intermediary conditions: with the spin that does not lie on the velocity vector at all, nor is it completely orthogonal with respect to the plane of motion.

The wave variation triggering the reaction of the **R.S.P.** in Lorentz force propagates in space-time at the velocity of light as **a variation in wave energy**; this is considered as a photon-wave train that is physically perceivable and capable of inducing in other particles some variations in momentum, and producing other known optical effects.

When the electron reaches some relativistic velocities, this wave variation has a specific correspondence in the experience and can be identified with the synchrotron radiation.

In order to make the wave interpretation of the Lorentz force more effective and the situation clearer, we should look for a plausible reason of the wave nature of electric charge, by investigating with full particulars the behavior of the resonant wave-front which is in the vicinity of the resonance orbit.

We shall deal with this subject further on. Now, let us make a step-back and go on to analyze the explicative possibilities of the simple model of mass considered as a generic source of elementary waves.

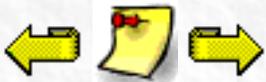

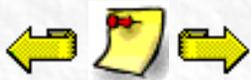

# The wave nature of inertia

The waves of the particle-wave source must be considered as having a spherical symmetry, and this is the physical reality of a wave field. Let us consider now only those portions of wave-fronts that propagating along the axis passing through the center of a field can be considered as plane and parallel to one another.

These waves, that are temporal perturbations of the lattice of the Schild's space-time, can be classically described as discrete sine waves that, being continually emitted by the source, must have a stationary description.

**24)**

$$\varepsilon = \cos(K_0 \, x') \cos(\omega_0 \, t') = u(x) \, \psi(t)$$

$$(\omega_0 = 2\pi; K_0 = \frac{2\pi}{\lambda_0})$$

where $\nu_0$ and $\lambda_0$ are the frequency and the wavelength of the source at rest.

If the source is in motion, we shall have two different frequencies due to Doppler effect: one in the direction of motion $\nu_1$ the other one in the opposite direction $\nu_2$.

In order to know what kind of wave will be observed from a real observer at rest, we have to apply Lorentz transformations:

**25)**

$$x' = \frac{x - vt}{\sqrt{1 - \beta^2}} \; ; \; t' = \frac{1}{\sqrt{1 - \beta^2}} \left( t - \frac{vx}{c} \right)$$

from which it follows that:

**26) 27)**

$$K_0 x' = Kx - \beta\omega t \quad \omega_0 t' = \omega t - \beta Kx$$

Consequently, the observed wave is:
**28)**

$$\varepsilon = \cos(Kx - \beta\omega t) \cos(\omega t - \beta Kx) = u(x,t) \, \psi(x,t)$$

As shown by Claude Elbaz (*), it is possible to take the wave state of particle into account and describe it as a stationary composition of plane waves in superposition of frequency $\nu_1$, $\nu_2$.

*(*) Claude Elbaz. C.R. Acad. Sc. Paris, t. 298, Série II, n¡ 13 1984.*

It is possible when we think of geometry, where for:

$$d\omega_1/\omega_1 \leq 1 \; ; \; d\omega_2/\omega_2 \leq 1 \; ; \; d\nu/\nu \leq 1$$

the phase:

**29)**

$$S = \omega t - \beta Kx$$

of the phase wave :

**30)**

$$\psi(x,t) = \cos(\omega t - \beta Kx)$$

fulfills the **Hamilton-Jacobi** equation

**31)**

$$\frac{1}{c^2}\left(\frac{\delta S}{\delta t}\right)^2 - \left(\hat{U}S\right)^2 - k_0^2 = 0$$

While in the field of diffraction, where:

**32)**

$$\Delta\omega \, \Delta t \cong 2\pi \ \text{è} \ \Delta K \, \Delta x \cong 2\pi$$

the phase wave $\psi(x,t)$ fulfills the **Klein-Gordon** equation:

**33)**

$$-\hat{U}^2 \Psi + \frac{1}{c^2}\frac{\delta^2 \Psi}{\delta t^2} + k_0^2 \Psi = 0$$

In the light of the Wave Field Theory, the Hamilton-Jacobi equation has a precise tangibly physical meaning from the wave viewpoint, that only now we can attribute to it.

It virtually identifies the trajectory of a photon, describing the place of the points of the space where the wave-fronts of the wave train-photon are parallel to one another.

**The Klein-Gordon equation would become more and more significant, if it were not falsified by the serious problem of the double solution we have to associate to it when it describes the De Broglie material waves.**

Indeed for:

**34)**

$$\omega^2 - p = m^2$$

the Klein-Gordon equation has two solutions: a positive solution describing a positive momentum and a negative one that leads us to consider the possibility of the existence of negative frequencies beside the positive ones.

**35)**

$$\frac{\delta^2}{\delta t^2}\Psi(x,t;p) - \hat{U}^2\Psi(x,t;p) = -m^2\Psi(x,t;p)$$

Till now only the positive solution has provided a physical meaning. But now, in the light of the Wave Field Theory, observing the wave situation in the vicinity of the particle-wave source, **we can verify the physical existence of positive and negative energy variations.**

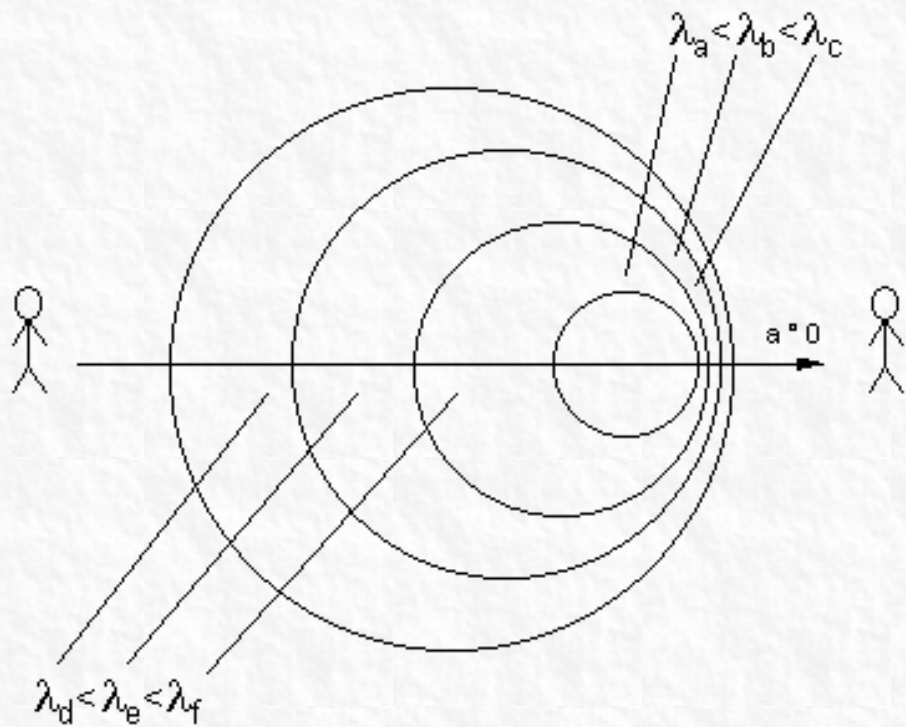

**FIG.7**

The change in the wave energy in the vicinity of the particle-wave source can be considered both a variation due to increment and a variation due to decrement.

Consequently, according to the considerations on the relative symmetry principle we have to call **"positive energy"** the variation due to increment and **"negative energy"** the variation due to decrement.

Both of them should influence the motion and rest state of the particle-wave source.

Both of them can intervene in the variation in momentum in the vicinity of the particle-wave source.

However, it is essential to say that regardless the type of variation, the relative symmetry principle must intervene in both cases to the same extent.

On this base, we are therefore obliged to look for the possible existence of a negative energy in physics.

## Inertia is the negative energy.

Following the Popperian philosophy, **let us try a falsification of the statement and propose an experiment.**

Let us speed a mass up suddenly, and verify the variation in the momentum of a testing body in the vicinity of a mass, in the side opposite to the direction of acceleration of the mass, regardless the active gravitational forces, that is, subtracting the effects due to gravity.

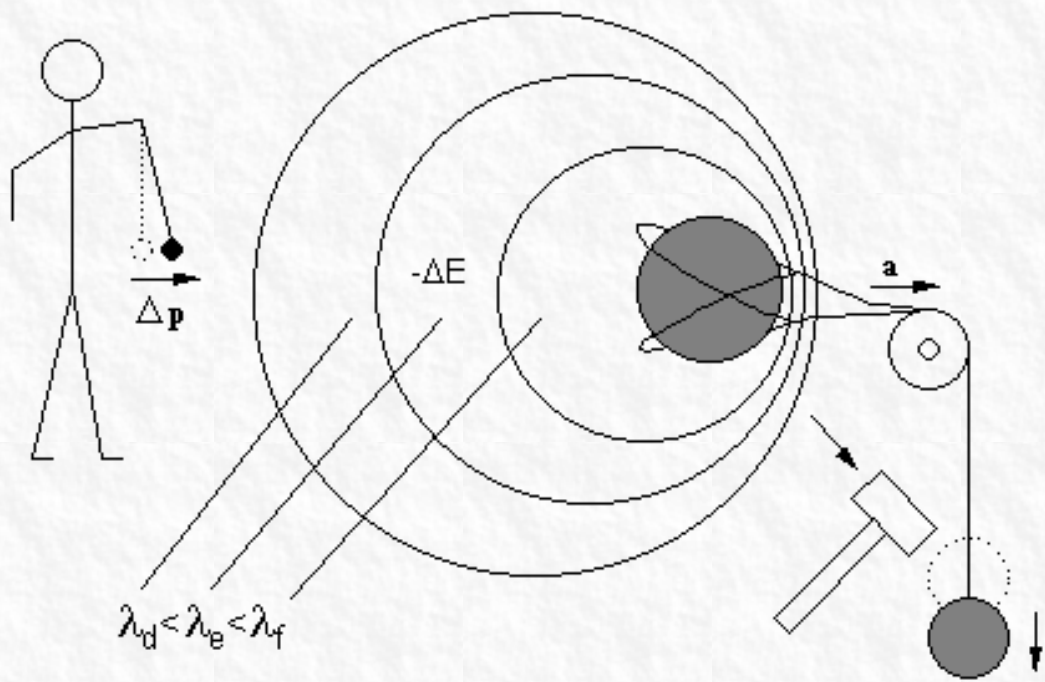

**FIG.8**

In the zone in which there is **-E**, a reduction of wavelength has occurred along with the creation of a wave energy **"hole"** owing to the acceleration of the mass. The relative symmetry principle pushes the nearest masses toward the hole.

It is as if in a constant-pressure atmospheric environment there were a loss in pressure, localized in a certain zone; all bodies are pushed toward that zone because of a general redistribution of pressure. The nearer the bodies to the trough are, the more they will undergo an energy action and remarkable effects.

Even the mass which is accelerated is subjected to the attraction of the negative energy "hole" **-ΔE**.

**This effect we could improperly impute to a force, is what we have till now called: inertia of mass.**

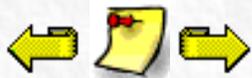

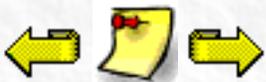

# THE WAVE NATURE OF GRAVITY

Let us examine a simple physical situation in which the base units of the **M.K.S.** system intervene in the same way as in the Cavendish experiment.

Let us consider two spherical bodies of masses $m_a$ and $m_b$ considered as totally isolated in a space lacking in significant fields, both weighing 1 kilogram and being at a distance of 1 meter from each other, .

Each body emits spherical waves of rest wavelength $\lambda_o$:

$$\lambda_0 = \frac{h}{c m_0}$$

Owing to the gravitational force the two isolated bodies would precipitate against each other, constantly increasing their approach velocity. In order to maintain the wave situation simple and the wavelength emitted in any direction in the vicinity of each body constant, we have to bind the bodies with a thread having an insignificant mass so as to oppose the tendency of the bodies of attracting each other with the tension of the thread by maintaining them still in space at the desired distance.

The important points of the gravitational wave interaction are all on the straight line passing through the centers of the two bodies.

Let us observe the behavior of the wave energy along this straight line in the time which elapses from the emission of a wave-front to the following one, from each body, dividing such an interval into various times and blocking the wave-fronts in such various times.

**The wave energy state in the space between the two bodies is variable.**

**This is the conclusive observation for the wave explanation of gravitation.**

If we take into account the additivity of masses after the experiment(*), we discover that the wave energy is variable in the space between one mass and another, as shown in **Fig. 9**.

As for masses having the same value, the energy in the point **A** is given by:
$E_{ab} = E_a + E_b = 2E$, as much as in the point **B** that is symmetrical of **A** with respect to the center of mass of the system. It follows that: $E_{ba} = E_b + E_a = 2E.$

Actually, the observable mass outside the system along the straight line passing through the centers is given by the sum of the masses of the two bodies.

**On the contrary, in the space separating the two bodies, the wave energy varies in time, oscillating in a cyclical sequence between the minimum value 1E and the maximum value 2E.**

(*) *A study on the nuclear mass defect will show this is not always true.*

In order to establish how it occurs, let us observe in **Fig. 9** the wave state, and let us try to block the wave-fronts in time $t_1$.

In the case **a)** at time $t_1$, the wave-fronts deriving from the mass ma are intercalated to the wave-fronts deriving from the mass **m<sub>b</sub>** both inside and outside the system.

In the case **b)** at time **t$_2$**, the wave-fronts propagate in opposite direction from each other and are superimposed to each other between the two bodies within the system, while they intercalate to each other outside it.

*We must not forget that the elementary waves we are speaking of cannot be described as sinusoidal functions, but they must be considered as perturbation bidimensional surfaces of the Schild lattice.*

*When they are in a geometrical superposition, the sum of two perturbations produces nothing but another perturbation.*

*Two waves intercalating their wave-fronts, simply produce a wave of double frequency, without having any interference phenomenon.*

- Between time **t$_1$** and time **t$_2$** the wave energy passing through the central zone, has changed from the value **2E** to the value **1E** due to decrement.

- Inside the system, the energy **-ΔE** has changed, creating a negative energy "hole" toward which the relative symmetry principle pushes the two masses.

The variation in the negative energy **-ΔE** is a sort of **stationary cyclical change** that we could call: negative wave or ***antiphoton.***(**)

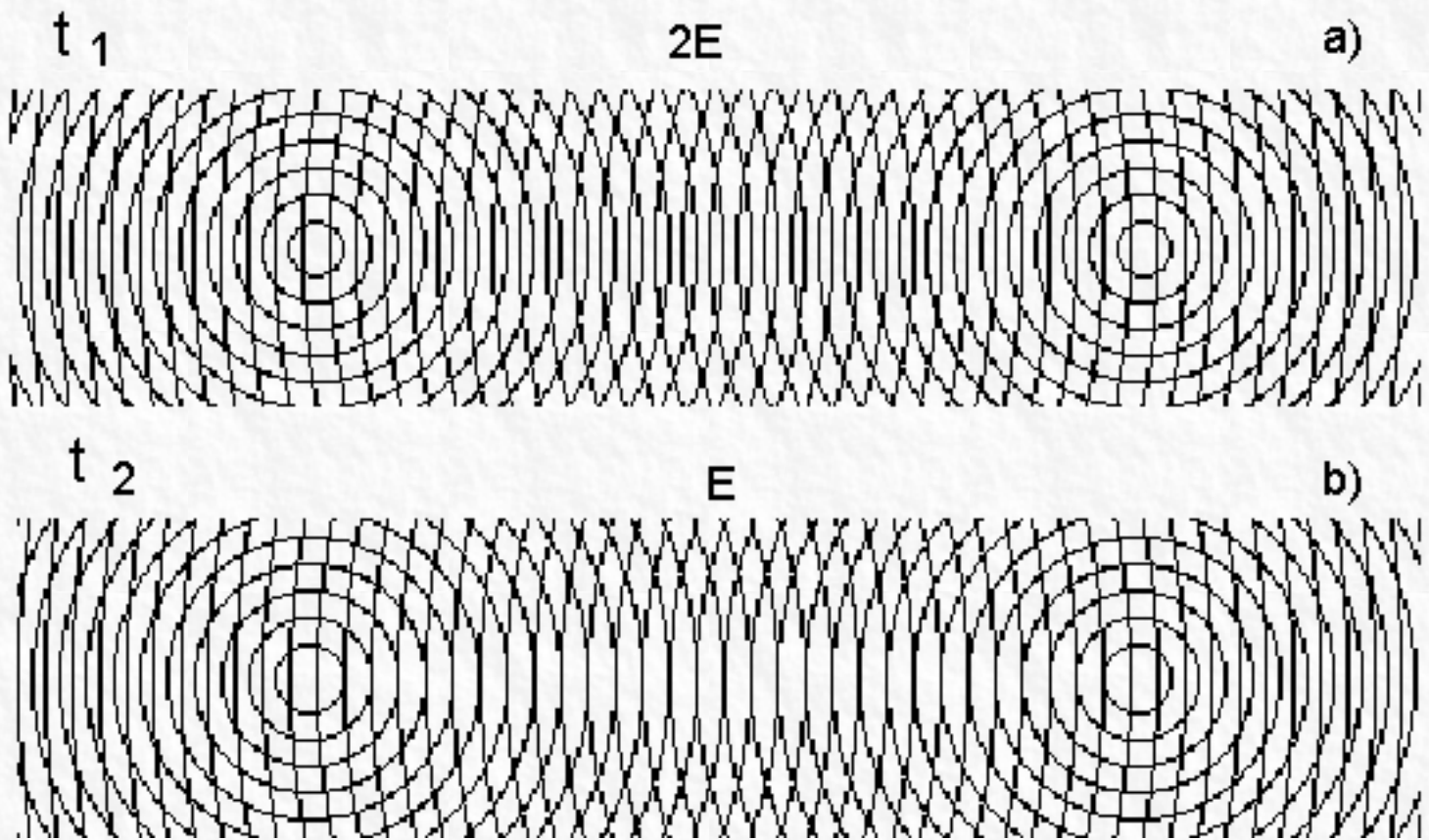

**FIG.9**

(*) *The amplitude of the elementary sine waves describes the value of the radius of curvature of a surface, and it never cancels out at all when it is superimposed (only in the mean point among the masses, the two waves cancel the radius of curvature).*

(**) *We could also call it graviton, realizing that it is not a particle but a wave train-negative photon of*

*frequency = 1/t<sub>g</sub>.*

The period $t_g$ of the negative wave must be considered as the time characteristic of the elementary wave action in the gravitational wave interaction.

**36)**

$$t_g = \frac{\lambda}{4c}$$

It is specific for each pair of bodies and it is linked to the value of the relativistic wavelength $\lambda_1$ of the elementary wave, which is emitted by the masses $m_a$ and $m_b$ due to Doppler effect, when the bodies move freely toward each other. However, it does not depend on their distance.

Actually, when we take away the threads maintaining the two bodies still at the desired distance, after a time $t_g$ the momentum of each body will be changed from zero to: **p=mv**.

Along the straight line passing through the centers, the velocity of waves **c** is constant, while the time of the energy-wave variation $t_g$ varies in the space among the bodies.

On the contrary, the ratio $t_g/\lambda_1$, whose constant value is: **1/4c**, remains unchanged during the approach of masses.

The first impulse for each body is given by the variation in momentum in time $t_g$. (*)

**37)**

$$p_1 - p_0 = \int_{p_0}^{p_1} dp = \int_{t_0}^{t_g} F_g \, dt$$

Then, considering $p_0=0$ e $t_0=0$ and the initial rest condition, we have for the mass $m_a$: $p_a = F_{ga} \, t_{gb}$ where time $t_{gb}$ is the period referring to the waves $\lambda_b$ coming from the mass $m_b$

**38)**

$$t_g = \frac{\lambda_{1b}}{4c}$$

the non-relativistic momentum of the mass $m_a$ is:

**39)**

$$p_a = m_a v$$

*(*)Let us discuss about integrals and differentials as if we were in a continuum, seeing that we consider the dimensional quanta of a discontinuous set very small. However, the question should be coped with after analyzing the finite differences.*

Therefore, the gravitational wave force, which does not depend on the distance among the masses yet, has a first component:

**40)**

$$F_g = \frac{4h\upsilon}{\lambda_a \lambda_b}$$

In order to find the second component, we have to take the hyper cubic lattice of the Schild space-time and its perturbations into account, considering the importance of the radius of curvature of the elementary waves establishing the effectiveness of the relative symmetry principle in the gravitational interaction.

Let us apply the formula **15)** for $\lambda_a = \lambda_b$ in order to obtain the velocity of the waves that are parallel to one another outside the system of the two bodies, and which will inversely depend on the Gaussian radius of curvature of the wave-fronts portions.

**41)**

$$v = \frac{Kc}{t^2 \sqrt{5}}$$

**K** is the factor of proportionality with dimensions **[L$^2$].**

Given the distance quantization and the following surface quantization, the unitary surface parameter that is proportional to the velocity **v** is: **K=N L$^2$**, where **N** is the number of the surface quanta **L$^2$** of the parallel "effective" wave-front.

When **K** is substituted in the formula of the velocity **v** , it follows that:

**42)**

$$v = \frac{NL^2c}{r^2 \sqrt{5}}$$

We obtain therefore the second factor that in the formula **F$_g$** gives :

**43)**

$$F_g = \frac{4hL^2c}{\sqrt{5}} N \frac{1}{r^2 \lambda_a \lambda_b}$$

There are two unknowns in this formula: the discrete elementary length **L**, and the number **N** of the discrete elementary surfaces forming the effective wave-fronts, which constitute the wave trains activating the relative symmetry principle.

Let us formulate the "ad hoc" hypothesis, we shall try to justify in the next chapter,*( partly depending on considerations on the symmetry of the domains of natural constants)* that a "reasonable" value of the discrete elementary length depends on the absolute magnitude of a terminal mass.

**44a)**

$$m = \frac{h}{\lambda c}$$

for the unitary $\lambda$

**44b)**

$$L = \lambda_L = \frac{h}{mc} = 4{,}884356 \cdot 10^{-84} \, m$$

We put forward a hypothesis (we will estimate afterwards its soundness) on the value of the discrete length equivalent to the dimensional quantum **L**, through which we can calculate the value of the number **N** of the effective wave-fronts in the wave formula of gravitation.

When we take the unitary parameter **N** from the experimental findings of $F_g$, we shall obtain the following result:

**45)**

$$N = 1/L \cdot 1{,}8777557 \cdot 10^{14}$$

(where **1/L** is the pure number showing the number of the linear quanta **L** in 1 meter)

The number **N** could seem at first sight lacking in physical meaning, but if it is resolved into factors, it will became of fundamental importance.

**46)**

$$N = \frac{1}{L}\left(137{,}024 \cdot 100^5\right)^2 = \frac{1}{L}\left[\frac{100^5}{\dfrac{e^2}{4\pi\varepsilon_0\, \hbar c}}\right]$$

Where the number **137,024** being too similar to the inverse of fine-structure constant, is identified with it.

If such an identification were confirmed by other elements and considerations, we should have a first reasonable element so as to link elementary waves and their gravitational action to electromagnetic waves and to interactions among electric charges.

Afterwards, we will be able to discuss about sound subjects supporting this thesis. Now, we can already estimate the capacity of the model to describe gravitational interactions.

Through it we can follow the various quantum phases of the gravitational wave action, highlighting the **"whys"**, tightly connected to gravitation, apart from the **"wherefores"**.

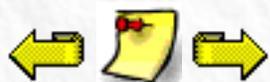

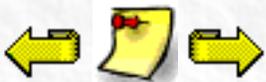

# Terminal velocity for masses

We know that the rest wavelength of an electron decreases owing to Doppler effect when the electron mass is accelerated at velocities approaching **c**:

**How long will its wavelength go on to decrease?**

As to the relativistic increase in mass, the question is as much as to wonder: till which value can a mass increase, if accelerated at velocities approaching the velocity of light?

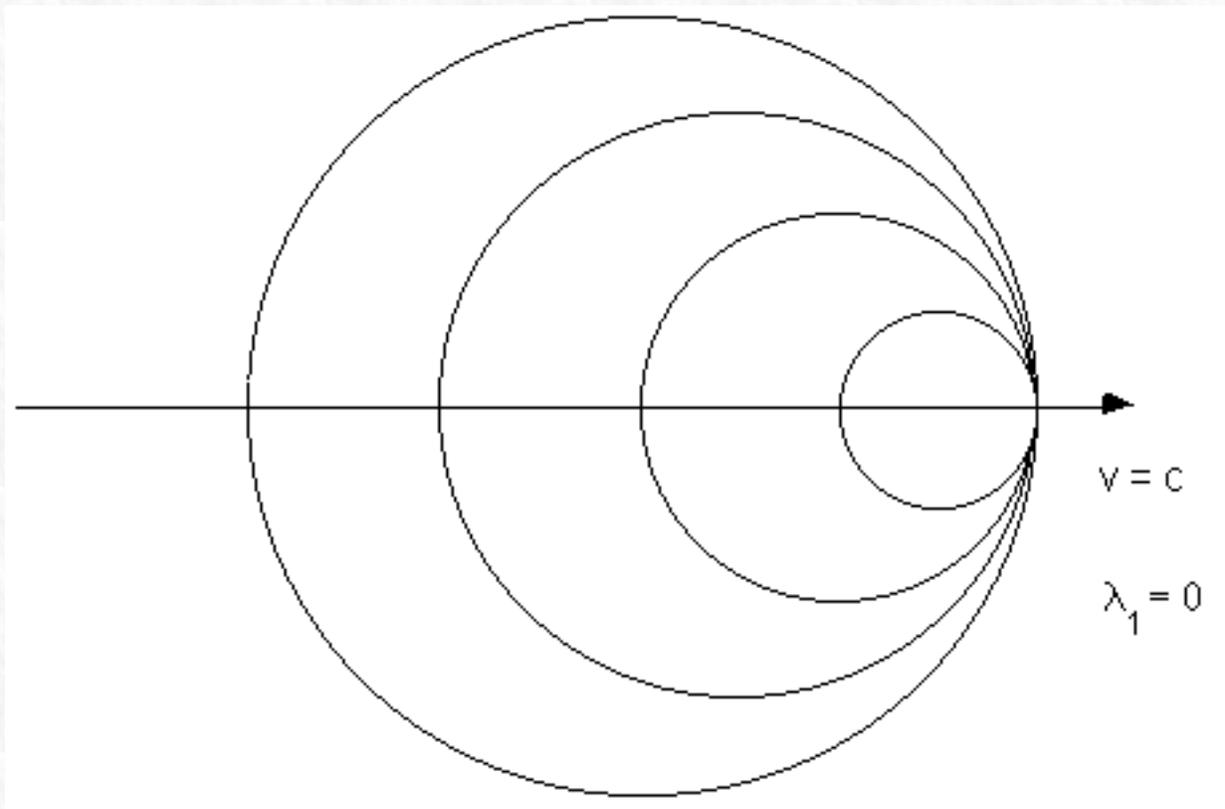

Could a mass ever reach the velocity of light?

That is as much as to wonder: is it possible for the wave source to reach the velocity of the waves it produces?

If it were possible, it would mean that the wave source produces waves whose wave-fronts are superimposed one upon another, and they would have therefore a null wavelength.

It cannot certainly occur in the Wave Field Theory. In such a theory it does not make any sense to speak of a mass having a null wavelength. The same goes for speaking of infinite mass in any other physical theory.

Nor does it make any sense to consider in the De Broglie wave theory any mass to which a wave train with a null wavelength is associated.

The questions find a common answer in the discrete space-time and in its length quantization.

Certainly, a wave deriving from a wave source-mass can never have a wavelength inferior to: $\lambda_{min} = L$

---

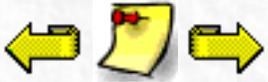

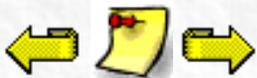

# The fifth antigravitational interaction

The electron at rest is the stable particle with the smallest mass existing in nature, while proton has a mass **1836** times greater than the first one.

Considering the limits of the velocities of masses, derived from the existence of a terminal length **L**, we deduce that proton has a terminal velocity necessarily slower than electron's.

Indeed, the proton's rest wavelength is less far from **L**; its velocity increases, its wavelength emitted in the direction of motion will first reach the terminal length L.

This means that its terminal velocity is certainly slower than the electron terminal velocity.

As we consider greater and greater masses emitting waves with smaller and smaller wavelength, we will have to verify the reason why their terminal velocity is slower and slower, and less close to the velocity of light.

If we go on to reason, we will deduce that a **"maximass"**, being so great as to have a rest wavelength equal to the terminal length **L**, will never have any velocity, and it will be obliged to be at rest forever.

At rest compared to what, or to whom?

It is known from relativity that an absolute frame of reference does not exist. It is therefore obvious that the rest of the maximass can exist only in comparison with other observers.

However, since the Doppler effect exists both when a wave source moves toward the observer and when the observer moves toward the wave source, no material real observer endowed with mass will ever have a velocity approaching the maximass.

This is not possible because the waves constituting its mass should add to the waves produced in the vicinity of the maximass, and they should intercalate their own wave-fronts to those of the maximass, producing a total wavelength inferior to the terminal wavelength: $\lambda_L = L = 4.88 \ 10^{-84}$ **m.**

- Since it is geometrically impossible, no mass can ever exist in the vicinity of the maximass.

- It follows that no mass can ever be attracted by it.

- Not only in the vicinity of the maximass is gravity null, but neither masses nor wave sources can ever exist in its vicinity, **(this agree with De Vaucouleur)**.

- It is the only existing isolated system.

$$F_{GW} = \frac{4hL^2cN}{\sqrt{5}r} \left[ \frac{1}{r\lambda_a\lambda_b} - \frac{\left(\dfrac{1}{\lambda_a} \cdot \dfrac{1}{\lambda_b}\right)\cos\phi}{r\sqrt{1 - \dfrac{L^2}{\lambda_a \ \lambda_b}}} \right]$$

**Fig. A1**

*This formula expresses the variation in the wave gravitational force, depending on the entities of masses. Expressing the prevalence either of attraction or repulsion with regard to the cosine function of the angle of view,*

*both from the distance separating the masses **a** and **b**, and **from the distance where the observer is**.*

We could dare to say that the observable universe is perhaps the maximass.
But if we verify numerically the real value foretold by the **WFT**, we shall see that the terminal wavelength is strangely close to the wavelength of the mass of a standard Galaxy, that is about **1 · 10⁴¹ kg**.

$$\frac{h}{\lambda_{min}c} = \frac{m\left(1 - \dfrac{v}{c}\cos\phi\right)}{\sqrt{1 - \dfrac{v^2}{c^2}}}$$

**Fig. A2**

*Let us describe the wave mass depending on the maximum wave number, that is inverse of the minimum wavelength, multiplied by the ratio **h/c**, and consider such a maximum mass (maximass) responsible for its absolute angular distribution, or rather depending on its concentration and on its relative velocity with respect to the observer.*

When we take masses smaller than the maximass into account, we think that it can attract small masses, but only with an extremely weak force. However, such a mass would attract with a greater force small masses rather than great ones.

At least, such an effect of diversified gravitational force should exist in any case, for any mass, and also for the earth's mass on which there would be therefore a difference in the gravitational attraction of masses of different value: the greater the mass is, the greater the repulsive effect opposed to the gravitational attractive effect is.

**This effect could be easily attributed to a new repulsive force, a fifth interaction, whose existence would involve a confutation of the Galileo's principle which affirms that all masses fall in the vacuum of a gravitational field with the same acceleration.**

- An attractive difference of masses with the same value but with a different radius or a different spatial distribution would occur as a further effect.

- Consequently, passing to more general considerations, the terminal wavelength of a macroscopic body composed by several elementary wave sources would be connected to the value of the effective wave-front of the waves totally deriving from the body.

Moreover, this "*effective*" wave-front would be a function of the parallelism of the waves deriving from the body and would depend on the spatial distribution of the elementary sources constituting it, and consequently on the "*form*" of the body.

- Bodies of equal mass and density but different in forms can have different terminal velocities.

- A disk moving in vacuum in the direction of its axis can attain with the same energy a velocity speeder than a sphere with the same mass.

*(And probably this will make ufologists happy).*

- As a logic inevitable consequence, a disk in vacuum can be accelerated more easily in the direction of its axis, rather than in the direction of its diameter.

- A pole will absorb greater energy for attaining a certain velocity in vacuum when its axis is turned in the direction of its velocity, rather than when its axis is perpendicular to it.

- Materials with the same mass but with a different molecular structure or a different nuclear structure would have different weights, since they would be attracted in different ways.

At cosmic level, material real observers should verify that all masses have rates of departure from masses close to the terminal mass. In that case, indeed, the repulsive force can overcome the attractive gravitational force.

Besides, the farther the observer is, the speeder the rates of departure are, seeing that widening the field of observation, he includes a greater number of masses in his observation.

**All of that easily leads us to imagine that in the universe the black holes and the stars of neutrons would have a little probability of existing.**

Naturally, that upsets the framework till now introduced by astrophysicists and cosmologists, and destroys dozen years of research on the first micro - nano - pico seconds from the birth of the mythical big bang, making useless many researchers' studies of a lifetime on the hypothetical structure of the products of the extreme matter concentration: the black holes

Anyway, an astrophysicist or cosmologist, whose success has been achieved after many years of academic relations, lectures, articles in specialized magazines, successful books, lessons and seminars, is unlikely to deny a whole life devoted to the search for a non-existent ghost.

> BY THE WAY, WE HAVE TO LAUNCH AN APPEAL TO A PERSONALITY WHO HAS BEEN WELL-KNOWN FOR A LONG TIME, CONSIDERED AS THE MESSIAH OF THE STANDARD MODEL COSMOLOGY:
>
> WE HAVE TO SAY SOMETHING TO **STEPHEN HAWKING:** WE ARE SORRY FOR YOU, WHO HAVE SURELY WON THE BET WARDING OFF ILL-LUCK MADE IN 1975 WITH YOUR FRIEND **KIP THORNE** TO MAKE SURE THAT, AFTER ALL, THE BLACK HOLES WERE NOT AN HOLE IN THE WATER.
>
> MAKE HIM PAY YOU THEREFORE A FOUR YEARS' SUBSCRIPTION TO THE "PRIVATE EYE" AS A CONSOLATION PRIZE FOR HAVING DEVOTED YOUR LIFE TO "WHAT DOES NOT EXIST" .
>
> NO ASTRONOMER CAN EVER IDENTIFY THE EXISTENCE OF A BLACK HOLE IN **CYGNUS X-1** FOR THE FACT THAT THE BLACK HOLES CANNOT EXIST. *

* (Taken from: "Albert aveva ragione - Dio non gioca a dadi". *Ed. Demetra  1996*).

And in order to justify the expanding universe, it would be no longer necessary to resort to the Big Bang hypothesis that is unlikely from the relativistic viewpoint.

**E. Fischbach**, in **1985**, found systematic deviations of the gravitational effects measured by **Eötvos** on materials with the same mass but with different nature. These deviations seemed to coincide with the gravitational anomalies verified in real orbits followed by artificial satellites, considerably different from those calculated according to the gravitational law.

**P. Tiebeger** of the university of **Seattle**, on the occasion of the annual School of Cosmology of **1990** informed *Varenna* of performing new experiments: he had immersed a copper sphere into water whose weight was equivalent to that of the volume of water displaced.

He demonstrated the existence of a repulsive, antigravitational effect acting in a different way in copper and water, when the tank got closer to a rocky mass placed orthogonally to the vertical one.

Afterwards, **P. Boynton**, astronomer at the university of *Washington*, verified the existence of some **antigravitational** effects through a torsional pendulum composed by a **toroid** constituted by two different materials (*aluminum and beryllium*).

The period of a pendulum, firstly well verified as a constant, changed when it got closer to a rocky wall, depending on which of the two materials was closer to the wall.

In **1995**, the daily newspaper "*L'Unità*" reported the news of an experiment for verifying the existence of a 5° repulsive antigravitational force. Professor **Focardi**, physicist at the University of *Bologna*, (one of the three discoverers of the *cold fusion* with Nickel announced in *Siena* a year and half ago) had performed qualified experiments in a tunnel under an artificial lake of **ENEL** (The Italian National Electricity Board).

These sound experiments verified the existence of an antigravitational force depending on the change in the mass of the water of lake coincident with changes in level in occasion of piloted fillings.

*("The Unified Field" that foresaw the existence of a fifth repulsive interaction was published in **1984**).*

- **(*Curiosity* – Two days before the lecture given by Focardi, Piantelli and Habel in Siena, in which they said not to understand the theoretical reasons of their results with Nickel at all, I was handing out my book "Albert was right: God doesn't play dice", in the bookcases of Siena, in which I displayed a - not ad hoc- sound and unitary theoretical explanation of cold fusion to advantage).**

- **(The Unified Field, that has been publishing since 1984, foresaw the existence of a nuclear force, not so much strong, depending on the fact that the model of the elementary particles of the W.F.T considered: the absence of an electrostatic repulsive force among the protons of a nucleus at the nuclear distance of 1 Fermi ).**

- **[Reprimand as for the open-mindedness of Professor Focardi]. After receiving both from me and from others a copy of "Albert was Right - God doesn't play dice", he decidedly denied any involvement in the new "Wave Field Theory " in his future researches.**

---

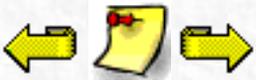

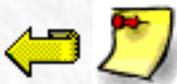

# The determination of "Uncertainty Principle"

In **1927** *Werner Heisenberg* subjected a strange development of quantum theory to the careful examination of the most important members of *Copenaghen School*. From the experiments, this development seemed to limit the possibility of knowing at the same time the parameters of position, energy or momentum of any particle described in quantum way.

After due considerations, *Bohr* and others found that the reason for the impossibility of knowing simultaneously the position and the energy of a particle was that **"the particle could not have any definite position and energy"**.

This statement, which turned the experimental impossibility into a philosophical principle, soon started to have a great importance in the description of microphysical properties, and became a physical principle that Sir *Eddington* named: ***"uncertainty principle''***. Eddington said:

*"All scientific authorities agree with the fact that at the bases of all physical phenomena there is the mysterious formula:*

$$\mathbf{q\ p - p\ q = ih/2\pi}$$

*We do not understand it yet. Perhaps, if we could understand it, we would not find it so fundamental.*

*A skilled mathematician has the advantage of using it (...) Not only does it lead us to the phenomena described by the previous quantum theories, like the rule of **h**, but also to many other phenomena. Therefore, the old formulas are quite useless."*

In the second member of the formula, besides **h** (acting atom) and the purely numeric factor **2$\pi$**, a mysterious **i** (square root of **–1** ) appears.

But it is nothing but a very famous subterfuge; in the past century many physicists and engineers knew very well that the square root of **-1** in their formulas was a sort of signal warning from waves or oscillations.

The second member has nothing strange; on the contrary, it is the first one which astonishes us.

We call **p** e **q** some coordinates and *momenta* consulting our dictionary of space-time world and our rough experience. But all of that does not explain us either their real nature or the reason why **p·q** differs from **q·p**.

It is obvious that **p** and **q** cannot represent simple numerical measures because if it were possible, **q·p-p·q** would be = **0**.

According to Schrodinger, **p** is an *"operator"*. Its *"momentum"* is not a quantity, but a signal which enables us to do a particular mathematical operation on some quantities following it.

It is obvious from *Eddington*'s considerations that, since the beginning the uncertainty principle has always been **"indeterminate"**. Afterwards, following a *Condon*'s idea, *Robertson* found at last an intuitive interpretation of the quantities **q** and **p**.

He got closer to the physical meaning of *"the uncertainty principle"*, finding that in order to establish with precision the position of a particle and to know simultaneously its momentum, there is a relation according to which the product of a probable mistake of positions and momentum is at least as great as Planck's constant

divided by **4π.** So, in general this relation should be valid:

$$\triangle \mathbf{p} \cdot \mathbf{q} \geq h/4\pi$$

Where Delta ($\triangle$) associated with **p** and **q** establishes that we have to consider a range of values deriving from a probable error in a **"mensuration".**

This is the only relationship with reality to be conceived for the uncertainty principle. It had however a great influence in asserting its supremacy in the philosophical relationship of quantum mechanics with the realist theories contesting its completeness.

Unlike quantum mechanics, no realist theory has ever given a theoretical justification to the uncertainty principle.

**Now not only is the Wave Field Theory able to provide a theoretical justification, but also a plausible model, entirely consistent with the wave model of particle as the *"spherical involute"*.**

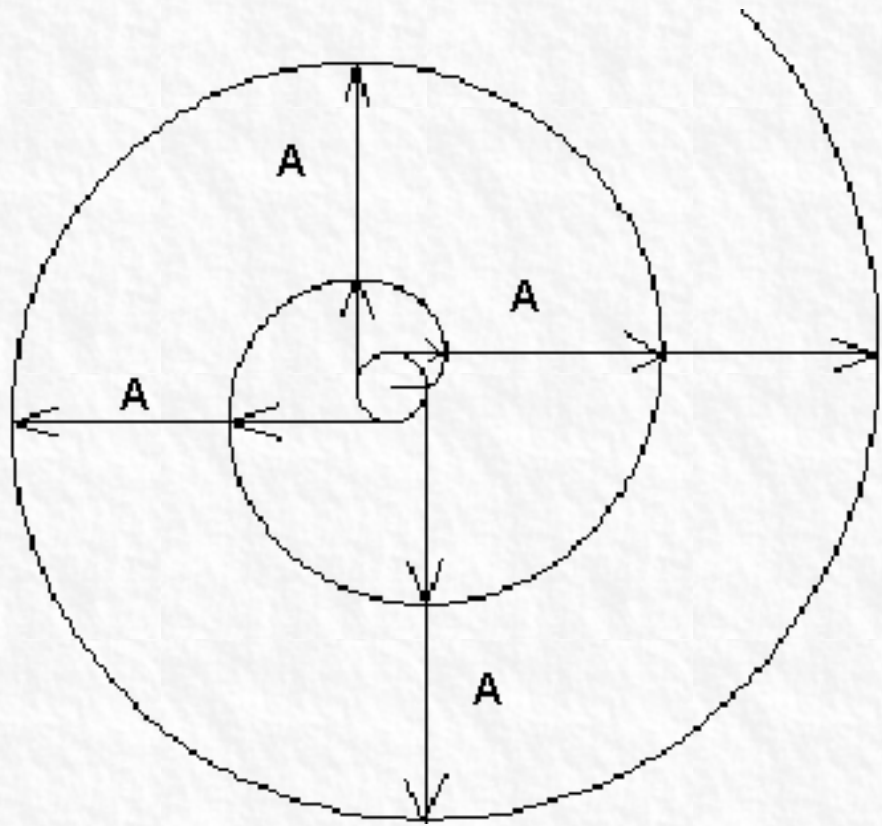

By observing a plane involute we realize that the wave structure is not a central symmetrical structure. Infact, in order to consider it this way, we must go away from the resonance orbit. Moreover, the more we go away from it, the more we can consider it as a spherical, symmetrical structure. While, in the vicinity of the resonance orbit, the involute has a precise eccentricity.

The static figure shows a specific point in the vicinity of the orbit in which the eccentricity, which lies near the resonant wave-front developing the involute, is highlighted. But in the real wave source, this eccentricy, or *"non-symmetry"*, moves at the velocity of light in the resonance orbit, with the resonant wave-front in the orbit.

**As a result, no mensuration can ever identify the real position of eccentricy, and therefore, nobody will ever be able to indicate or to place in space-time the center of the resonance orbit of a particle.**

This forced ignorance prevents us from knowing the exact effects of interaction between photons and particles.

**As a consequence, we cannot establish exactly the effect of the relative symmetry principle but with a sort of uncertainty that is a consequence of the wave nature and of the structure of the physical-geometric model of particle.**

All of that does not change the central position of the resonance orbit.

**This uncertainty, out of ignorance, does not give any philosophical evidence for the existence of real parameters concerning the central position of the resonance orbit of the particle.**

Through the wave model of the spherical involute we can interpretate the elementary particles' properties, and know the most mysterious mathematical theories of *Dirac* and *Schrodinger*, to which a consistent and causal physical interpretation has always been difficult associate.

*Schrodinger* interpreted the *Dirac* equation in order to establish the correct value of Spin and of the electron's gyro magnetic ratio.

The solution of *Dirac* equation for a free electron, for which there was already a mathematical justification of its spin, implies the existence of a special motion for particle which *Schrodinger* called: **Zitterbewegung**, a sort of trembling which *seems moving the particle by vibration at the velocity of light about the position it occupied in any point of its trajectory.*

In quantum electrodynamics this motion is not considered in a realistic way, and it is mysteriously linked to the existence of the *Heisenberg* uncertainty principle, seeing that it is hypotetically justified, in a non-causal way by the uncertainty of the position of an electron in a region having the same dimensions as Compton length.

On the contrary, now with an understandable and rational model, we can verify that the apparent motion is not real, but in a perfect casual and deterministic way it depends on the variation in eccentricity of the wave field traveling at the velocity of light together with the resonant wave-front on the resonance orbit of the spherical involute.

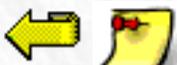

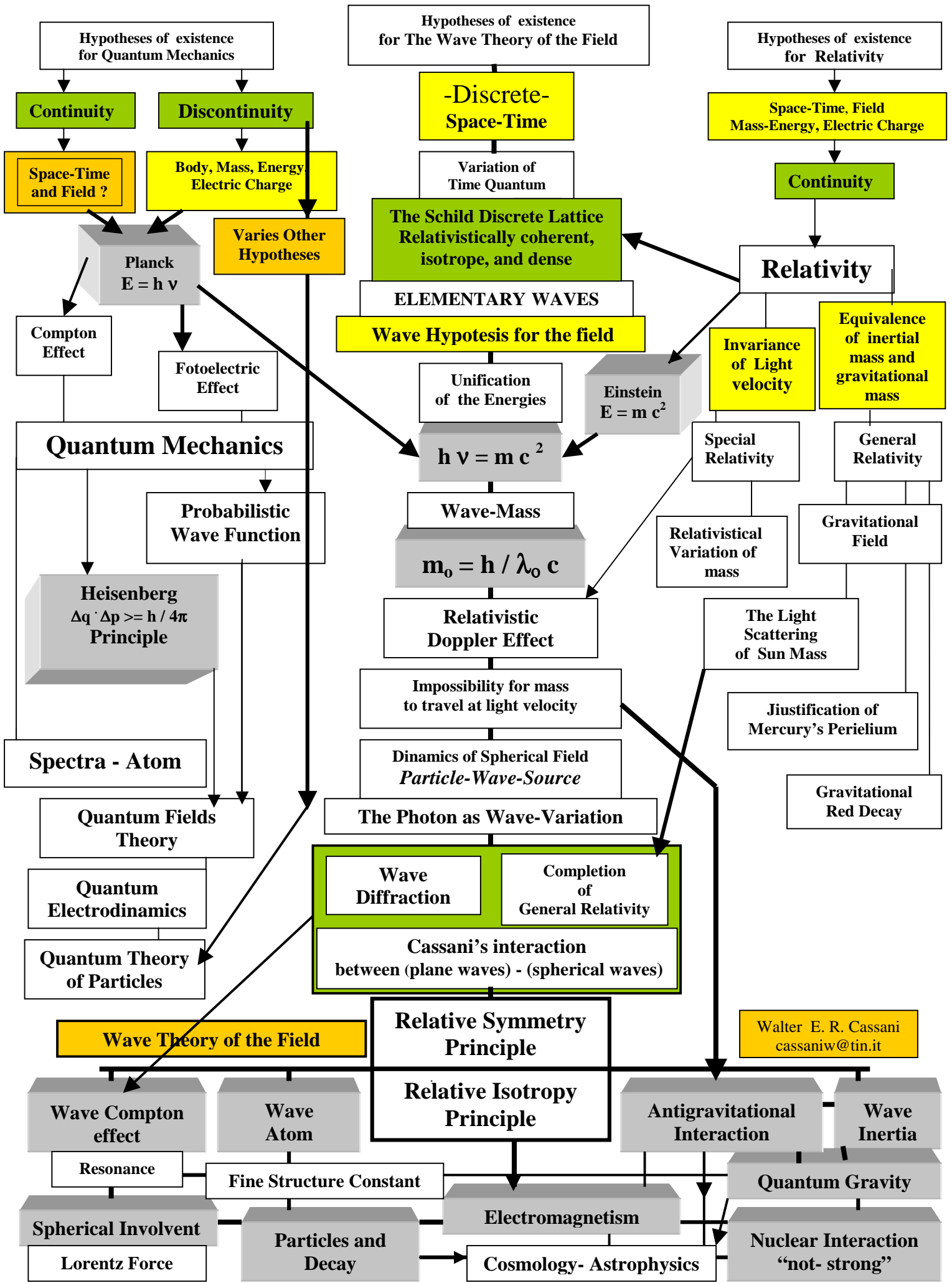

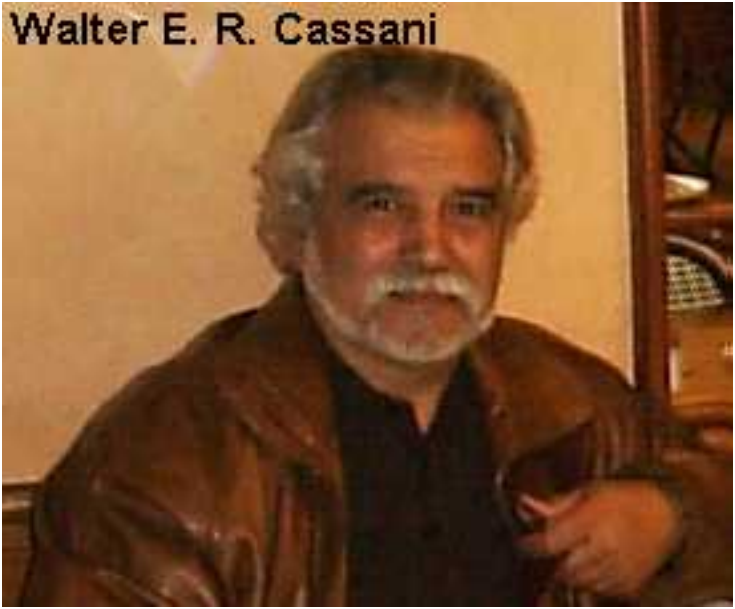